\documentclass[preprint]{acmart}
\AtBeginDocument{%
  \providecommand\BibTeX{{%
    \normalfont B\kern-0.5em{\scshape i\kern-0.25em b}\kern-0.8em\TeX}}}

\setcopyright{none}

\acmJournal{HEALTH}
\acmVolume{XX}
\acmNumber{Y}
\acmArticle{Z}
\acmMonth{8}

\begin{document}

\title{A Survey of Challenges and Opportunities in Sensing and Analytics for Cardiovascular Disorders}

\author{Nathan C. Hurley}

\orcid{0000-0003-3055-8825}
\affiliation{%
  \institution{Texas A\&M University}
  \department{Computer Science \& Engineering}
  \streetaddress{3112 TAMU}
  \city{College Station}
  \state{TX}
  \postcode{77843-3112}
  \country{USA}
}
\email{natech@tamu.edu}

\author{Erica S. Spatz}
\affiliation{%
  \institution{Yale University}
  \department{Center for Outcomes Research and Evaluation}
  \streetaddress{1 Church Street, Suite 200}
  \city{New Haven}
  \state{CT}
  \postcode{06510}
  \country{USA}
}
\email{erica.spatz@yale.edu}
\author{Harlan M. Krumholz}
\affiliation{%
  \institution{Yale University}
  \department{Center for Outcomes Research and Evaluation}
  \streetaddress{1 Church Street, Suite 200}
  \city{New Haven}
  \state{CT}
  \postcode{06510}
  \country{USA}
}
\email{harlan.krumholz@yale.edu}
\author{Roozbeh Jafari}
\affiliation{%
  \institution{Texas A\&M University}
  \department{Biomedical Engineering}
  \department{Computer Science \& Engineering}
  \department{Electrical \& Computer Engineering}
  \streetaddress{3112 TAMU}
  \city{College Station}
  \state{TX}
  \postcode{77843-3112}
  \country{USA}
}
\email{rjafari@tamu.edu}
\author{Bobak J. Mortazavi}
\orcid{0000-0002-2655-2095}
\affiliation{%
  \institution{Texas A\&M University}
  \department{Computer Science \& Engineering}
  \streetaddress{3112 TAMU}
  \city{College Station}
  \state{TX}
  \postcode{77843-3112}
  \country{USA}
}
\email{bobakm@tamu.edu}


\renewcommand{\shortauthors}{Hurley et al.}

\begin{abstract}
  Cardiovascular disorders account for nearly 1 in 3 deaths in the United States. Care for these disorders are often determined during visits to acute care facilities, such as hospitals. While the length of stay in these settings represents just a small proportion of patients' lives, they account for a disproportionately large amount of decision making. To overcome this bias towards data from acute care settings, there is a need for longitudinal monitoring in patients with cardiovascular disorders.  Longitudinal monitoring can provide a more comprehensive picture of patient health, allowing for more informed decision making.  This work surveys the current field of sensing technologies and machine learning analytics that exist in the field of remote monitoring for cardiovascular disorders. We highlight three primary needs in the design of new smart health technologies: 1) the need for sensing technology that can track longitudinal trends in signs and symptoms of the cardiovascular disorder despite potentially infrequent, noisy, or missing data measurements; 2) the need for new analytic techniques that model data captured in a longitudinal, continual fashion to aid in the development of new risk prediction techniques and in tracking disease progression; and 3) the need for machine learning techniques that are personalized and interpretable, allowing for advancements in shared clinical decision making. We highlight these needs based upon the current state-of-the-art in smart health technologies and analytics and discuss the ample opportunities that exist in addressing all three needs in the development of smart health technologies and analytics applied to the field of cardiovascular disorders and care.
\end{abstract}



\maketitle

\section{Introduction}
Cardiovascular diseases are the worldwide leading cause of death \cite{RN1}.  In 2016, cardiovascular diseases accounted for nearly 1 in 3 deaths in the United States. Real-time vital monitoring allows for care providers to track patient progress and to rapidly respond to any changes in patient condition. In the hospital, monitoring patients is part of routine clinical practice. Providers are able to monitor cardiac status and basic vitals from anywhere in the hospital at any time. Slight deteriorations in health can be observed and interventions put into place before patients suffer worsening harm.
However, length of stay in these acute care settings is often quite short \cite{RN69,  RN70}, representing only a small portion of a patient's life.  
Monitoring physiologic parameters and symptoms outside of the hospital setting can enable better detection and response systems before a person becomes acutely ill and requires hospitalization or after hospitalization to prevent early readmission to the hospital; however, many of the devices today are targeted to healthy people.
With the prevalence and ubiquitous nature of remote and wearable sensors, opportunities exist to broaden the applications of sensing and for adapting analytic techniques to enhance diagnosis, monitoring, and treatment among people with cardiovascular disease.

A challenge in monitoring patients with cardiovascular disorders is designing the technology and algorithms to support a variety of conditions and signs/symptoms, including heart failure (HF), coronary artery disease (CAD), and stroke, along with risk factors for these conditions such as hypertension (HTN). HF is typically a chronic condition where the heart is unable to drive blood forward through the body sufficiently or can only do so under damagingly high pressures. HF is a debilitating disease that causes significant global disease burden. In 2016, HF was the most rapidly growing cardiovascular condition in the world \cite{RN9}. CAD occurs when blood flow through the coronary arteries, the small arteries that provide blood to the heart, becomes impeded. This occurs both gradually as plaque builds up within the coronary arteries and suddenly when a plaque ruptures and clots. The former causes chest pain and exercise intolerance, while the latter, commonly known as a heart attack, can cause severe pain, loss of consciousness, and death. Each year around 800,000 Americans suffer a heart attack, and rapid care following a heart attack is a chief predictor for minimizing long term morbidity and mortality \cite{RN5, RN3, RN6}. Stroke is any disease impacting the blood vessels to the brain. In particular, acute stroke is a condition that occurs when either a blood vessel in the brain ruptures, or when one of those blood vessels becomes blocked. Stroke manifests with the sudden onset of neurological deficits, some of which may be irreversible. Stroke is the fifth leading cause of death in the United States and is a leading cause of long-term disability \cite{RN1}. HTN is a disease in which the arterial blood pressure (BP) is chronically elevated.  This elevation may remain asymptomatic for years but can ultimately lead to damage of multiple organ systems.  HTN is a very common condition, with middle-aged Americans having approximately 90\% lifetime risk of developing HTN \cite{RN65}.

Patients with cardiovascular disorders present a number of challenges for remote monitoring and diagnosis because of complexities within the diseases. Many of these diseases involve seemingly trivial symptoms that may suddenly change from a minor inconvenience to a debilitating lack of function. A patient with a given disease may feel well for multiple years, and then suddenly decompensate and require emergent care. Ideally, remote monitoring should be able to track the slow, daily progression of a disease and alert the patient and healthcare providers to worsening disease before decompensation and patient suffering. However, preliminary studies in remote monitoring have failed at preventing adverse events, such as in preventing repeated hospital admissions in patients diagnosed with HF. For example, the Telemonitoring in Patients with Heart Failure trial (Tele-HF) used patient self-reports of daily changes in symptoms, weight, and a variety of other factors (e.g., medication changes, depression scores, etc.) to identify worsening symptoms in an effort to intervene prior to another acute event, but did not find a statistically significant difference between control and intervention arms \cite{RN71}.
However, an analysis of participant subgroups did find that patient self-reported data could improve prediction of readmission likelihood, showing potential for more advanced analytic techniques to better identify participant risk and to improve estimates in this space \cite{RN34}.
A further exploration in automating the capture of relevant biometric signals, including heart rate, blood pressure, and weight, was similarly unable to find a statistically significant difference in control and intervention arms \cite{RN33}, suggesting that further exploration of additional biomedical signals are needed.

Remote sensing technologies have increased in prevalence and have made personalized health data collection feasible. In human activity recognition (HAR), wearable sensors and inertial measurement units embedded within smartphones and smartwatches have enabled the tracking of detailed motions \cite{RN72, RN73}. Coupled with nearable sensors that capture motion via video, these sensing systems allow for the tracking of motions of healthy participants \cite{RN74} to tracking of disease state with custom-built sensors, such as smartshoes \cite{RN75}. The data provided by these wearable and remote sensors has more recently enabled advanced machine learning techniques to identify more complex patterns of motions, better understanding personalized behavior \cite{RN76, RN77}. Eventually, these techniques have emerged to personalize models of activity recognition to individual users, providing the most robust interpretation of activities of daily living per user \cite{RN78}, enabling feedback and the measurement of clinical outcomes \cite{RN79}. This progression from the development of new sensing modalities to the analytic techniques that detect patterns within the data and finally to personalization in tracking and disease progression modeling is a pathway that should be leveraged for more advanced clinical disorders.

The development of new sensors to measure signs and symptoms of cardiovascular disorders would ideally enable a similar progression for tracking of cardiovascular outcomes. These new sensors would be able to identify conditions that may not be apparent to patients or providers, such as different sounds from the heart, slowly decreasing patterns of activity, or combination of vitals that may appear normal in isolation but may be indicative of risk given a combination of values and certain patient contexts. By identifying dangerous signs before symptoms manifest, earlier interventions can lead to improved health outcomes. A variety of technologies and machine learning techniques to this purpose exist in condition-specific settings \cite{RN37, RN80} to varied success \cite{RN81, RN82, RN83}. Understanding the pathologies of the disorders is important in understanding the clinical needs and opportunities that exist in developing new wearable and remote sensors for diagnosis and treatment of a variety of cardiovascular conditions and using advanced analytic techniques that are enabled from the collection of new, comprehensive patient data. 

In this survey, we discuss different sensing modalities that have been or that could be applied to tracking cardiac health in remote settings; in addition, we consider the opportunities that advanced analytic strategies present with the acquisition of remote sensing data for continuous risk modeling. As cardiac pathologies manifest, they can also be indirectly observed through physical changes in the body, potentially measured by sensors on or around the body. These changes can be utilized to track patient health, to plan interventions to maximize patient wellness, and to decrease the overall impacts of the disease. One of the oldest technologies used for assessing cardiac health is the stethoscope. In the digital era, the electronic stethoscope is a varied group of technologies that incorporate a microphone in order to automate acoustic diagnose and facilitate remote monitoring \cite{RN11}. Other technologies, such as photoplethysmography and sphygmomanometry, allow for remote measurement of the characteristics of the blood flow including heart rate and blood pressure \cite{RN84}. Doppler radar can detect vital signs such as respiratory rate and heart rate \cite{RN59}. Electrical techniques such as electrocardiography (ECG) or other conduction studies such as Bioimpedance can give insights into the internal physiology of the heart \cite{RN85}. 

\begin{figure}[h]
  \centering
  \includegraphics[height=0.85\textheight]{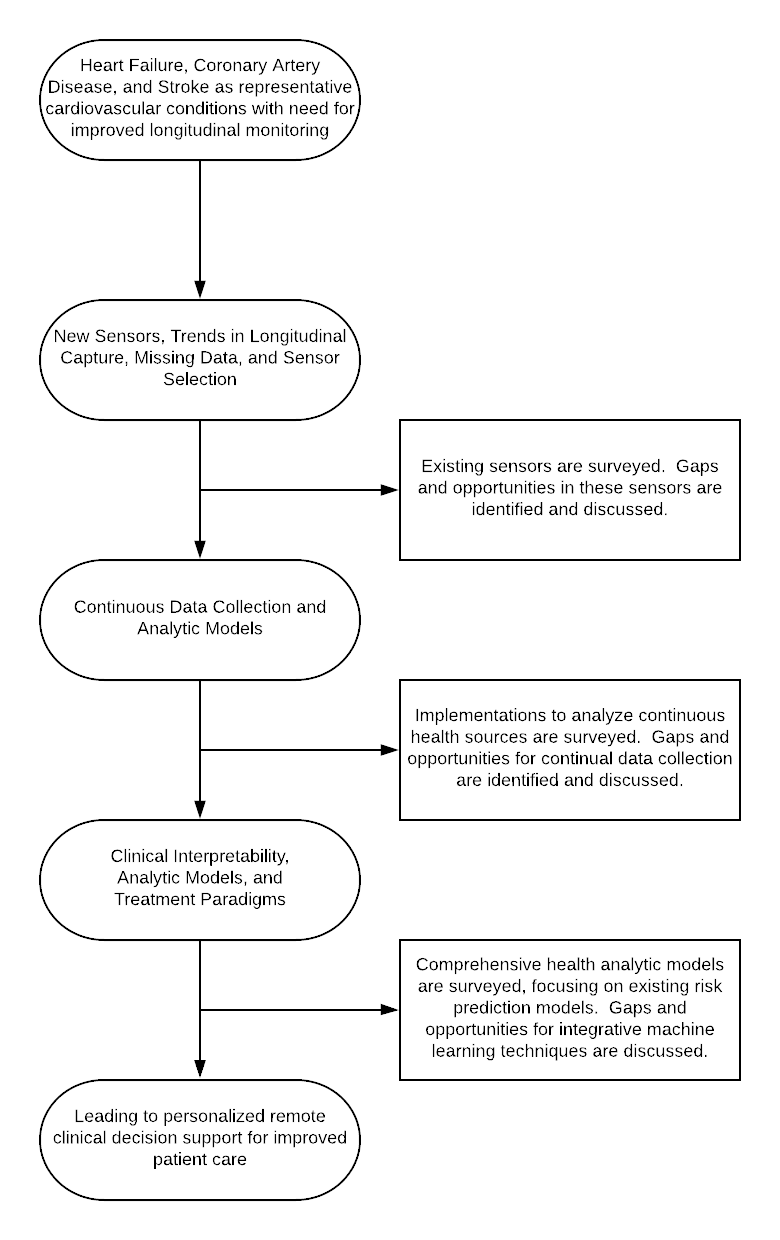}
  \caption{Overview of a workflow to developing personalized, remote clinical decision support tools for patients with cardiac disorders.  Needs are shown in three categories: needs in sensor development and data handling, needs in continuous data collection and analysis, and needs in developing comprehensive and personalized analytical models.  Addressing these three categories will allow for improved personalized remote clinical decision support for patients with cardiac disorders.}
  \Description{Overview of paper structure.}
  \label{fig:contents}
\end{figure}

Sensing systems provide for opportunities to proactively detect and alert patients and physicians to worsening health states.  However, to allow for timely and effective interventions as well as to rapidly evaluate the impact of those interventions, development of advanced signal processing and machine learning techniques need to keep pace with the development of raw sensor modalities. This paper presents a survey of state-of-the-art sensing technologies and analytics with respect to monitoring cardiovascular disorders, in order to highlight successes and provide areas for additional growth. Two key ways in which analytics associated with sensing systems can provide support are to monitor for the start of new disease and to track the progression of preexisting disease. Tracking the progression of existing disease is the easier task: once an underlying disease state is known, appropriate monitoring can be put into place and utilized to follow the progression of the disease. Monitoring for the start of new disease is more difficult, as the focus is more general. In either case, sensing and clinical characteristics can be combined for decision support with the aid of machine learning approaches. In this paper, we survey the current state of the art in patient monitoring and analytics for patient risk and care, highlighting needs and opportunities for advancements in the field of smart health with respect to monitoring signs, symptoms, and treatments in patients diagnosed with cardiovascular disorders.  The workflow described in this paper towards developing new tools for remote clinical decision support is shown in Figure \ref{fig:contents}.

\section{Case Studies and Needs}
This work considers three primary cardiovascular disorders for the review of gaps and opportunities, though by no means encompasses the entirety of technologies available for monitoring and treating these conditions nor the entirety of conditions to which these technologies could be applied. Instead, these conditions serve as meaningful examples in which technical solutions would be clinically impactful, following the same trajectory of development discussed in the prior section with regards to HAR. Those conditions are heart failure (HF), coronary artery disease (CAD), and stroke, with an additional emphasis on hypertension (HTN) as an important and common risk factor for these conditions. We briefly review these conditions and use them as case studies to discuss the current needs in smart health technologies.

\subsection{Clinical Conditions}

Heart failure (HF) occurs when one or both halves of the heart are unable to drive blood flow forward at the rate required by the body or can only do so under high pressures. This discussion of pathology will focus primarily on left-sided HF rather than right-sided HF, but the two are often closely associated and technologies for monitoring the two will have a large amount of overlap. The two will also often coexist.  HF can result from ineffective heart contractions, from high pressure limiting the effect of heart contractions, or from difficulty in filling the heart. The first two causes lead to HF with reduced ejection fraction (HFrEF), and the last leads to HF with preserved ejection fraction (HFpEF). Ineffective heart contractions can result from muscle damage caused by CAD, by chronic volume overload as seen in mitral regurgitation (MR) or aortic regurgitation (AR), or by a family of cardiac muscle disorders known as cardiomyopathies. High pressure can lead to HF either from aortic stenosis (AS) or from uncontrolled hypertension. In either case, the pressure that the heart has to work against is so high that the pumping becomes ineffective. Difficulty in filling the heart can be caused by ventricular hypertrophy, cardiomyopathy, fibrosis, disease around the outside of the heart (the pericardium), or by CAD. 

Coronary artery disease (CAD) is a family of diseases where blood flow through the small arteries of the heart, the coronary arteries, is restricted. This restriction can be caused by deposits of fatty plaques within the arteries, or by clotting caused by the rupture of one of these plaques. Depending on the extent of the blood flow restriction and the current oxygen demands of the heart, CAD may cause different symptoms. CAD is represented by a spectrum of conditions that are defined by specific clinical and physiological signs.

Stroke occurs when blood supply in and around the brain is acutely disrupted, and results in acute neurologic defects. Ischemic stroke is a type of stroke where a blockage in cerebral arteries rapidly blocks off blood flow, leading to cell death. Hemorrhagic stroke is a type of stroke where a blood vessel in the brain ruptures, rapidly raising pressure inside the skull and causing cell death. Transient ischemic attacks (TIAs) are similar in cause and presentation to strokes but resolve spontaneously. They are often an indicator of underlying disease and put the patient at increased risk for future TIA or stroke. The neurological pathology goes beyond the scope of this work, but there are several notable cardiovascular impairments that may cause a stroke. 

Hypertension is a condition where a patient's blood pressure is persistently elevated and is often a condition that serves as a modifiable precursor to each of the three cardiovascular disorders discussed \cite{RN86}.  Hypertension is divided by cause into two categories: primary (or essential) hypertension is hypertension without a particular medical cause, while secondary hypertension is hypertension caused by some other medical condition.  Primary hypertension accounts for roughly 90\% of all hypertension, while secondary hypertension accounts for the remaining 10\%.  Causes of secondary hypertension include renal disease and endocrine diseases that disrupt the body's natural control of blood pressure \cite{RN67}.  Essential hypertension is a diagnosis of exclusion, and requires ruling out the possibility of any secondary causes.  Risk factors for essential hypertension include both hereditary and environmental factors \cite{RN87}.  There is a strong association between hypertension, obesity, and insulin resistance.  Hypertension is associated with poor diet, excessive alcohol intake, and age. Additional discussion of normal cardiac physiology can be found in the Supplementary Appendix (Section \ref{sec:appendix}).

\subsection{Needs for Monitoring Cardiovascular Disorders}

\begin{figure}[ht!]
  \centering
  \includegraphics[width=\linewidth]{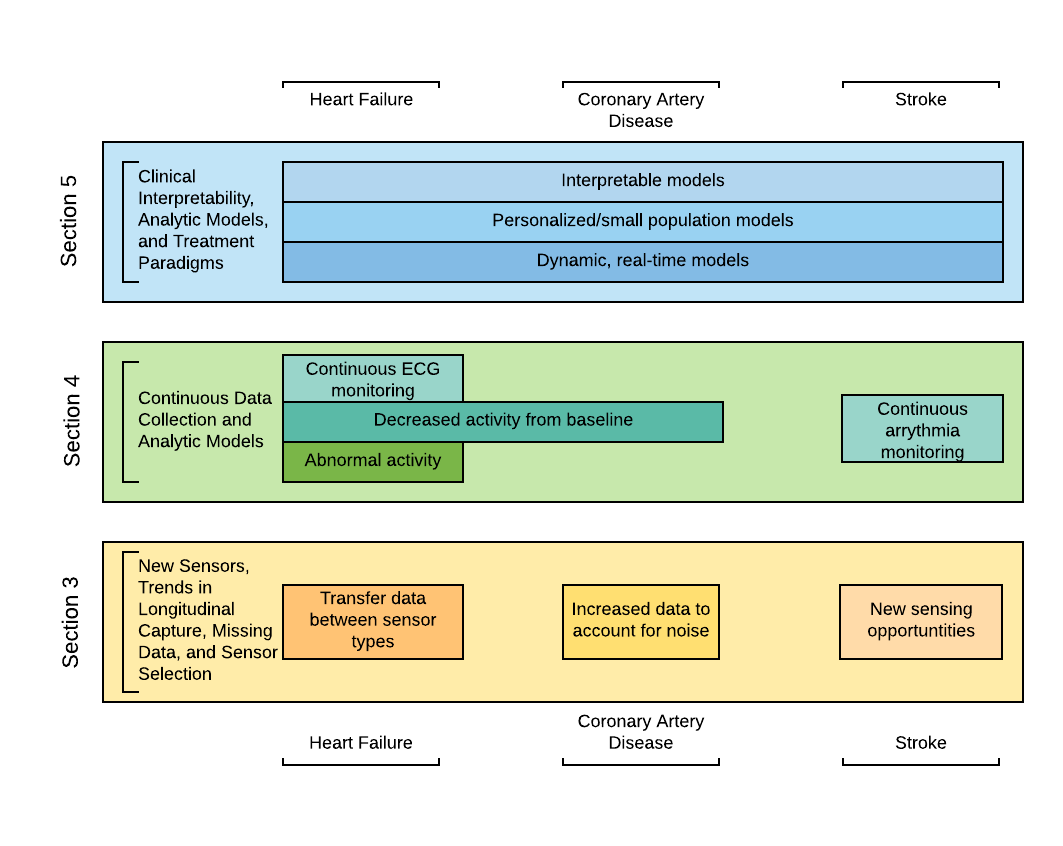}
  \caption{Progress from sensors to analytics (y axis) and how they relate to each of the three conditions (x axis).}
  \Description{Progress from sensors to analytics (y axis) and how they relate to each of the three conditions (x axis).}
  \label{fig:overview}
\end{figure}

Figure \ref{fig:overview} illustrates the three primary needs this survey will discuss: 1) the need for sensing technology that can track longitudinal trends in signs and symptoms of the cardiovascular disorder despite potentially infrequent, noisy, or missing data measurements; 2) the need for new analytic techniques that model data captured in a longitudinal, continual fashion to aid in the development of new risk prediction techniques and in tracking disease progression; and 3) the need for machine learning techniques that are personalized and interpretable, allowing for advancements in shared clinical decision making. A number of varied signs and symptoms exist for HF, CAD, and stroke.  The remainder of this section briefly introduces some common signs and symptoms.  Here, a symptom is a change caused by disease that is noticed by and likely an irritant to the patient, while a sign is a change that the patient may not notice or that may not be concerning to the patient.

In HF, the symptoms result both from insufficient blood flow and from excess fluid buildup. The three main symptoms that are associated with diagnosis of HF and quantification of its severity are dyspnea (shortness of breath) on exertion; sudden, choking dyspnea at night; and difficulty breathing while lying down. In left-sided HF pulmonary vein pressure increases, causing buildup of fluid in the lungs (pulmonary edema) that worsens while lying down. In right-sided HF systemic venous congestion results in fluid buildup in the periphery (peripheral edema) that worsens while upright, resulting in noticeable swelling in the wrists and ankles. HF is difficult to precisely define as it is a clinical syndrome resulting from many different heart conditions, and many variants exist. Therefore, attempts to understand HF and to monitor its progression must focus on identifying the symptoms and identifying cardiac dysfunctions. Symptoms that can be measured include peripheral edema (swelling of ankles, rapid weight gain), decreased activity, and changes in respiratory patterns when lying down versus remaining upright. Changes in blood flow to the kidneys result in decreased urine production during the day, and increased urine production at night. Patients with HF will therefore often get up frequently in the night. These patients will also likely change posture in the night, with patients with advanced HF needing to sleep upright. One of the most used classification schemes for HF is the New York Heart Association (NYHA) Functional Classification \cite{RN7}. In this classification scheme, classes are separated based on the physical activity that the patient is able to achieve and the discomfort that results from physical activity. Class I is when no symptoms are present, and in Class IV the patient is unable to perform any physical activity without discomfort and symptoms of heart failure are never alleviated. As can be seen, a variety of sensing modalities could be employed to track signs and symptoms of HF, from measurements of peripheral blood flow, respiration rate, exercise capacity, and posture while sleeping. This illustrates the need for new sensors that can measure each of these various symptoms. However, not every sensor may be worn at all times, due to excessive burden on the user. Therefore, there is a need for new sensing modalities that can track different patterns and trends in captured data, as well as transfer learning techniques that can be adapted to estimate values of sensors that may be malfunctioning or not worn. 

If the right set of sensors are selected and are designed to be worn longitudinally, new patterns and trends in signs and symptoms might be detected. In CAD, for example, restrictions in blood flow of the coronary artery may result in a condition called stable angina (SA). The rate at which the restrictions in blood flow occur, however, might change as the disease progresses. At some point, the restriction responsible for SA may rapidly increase, producing a situation where the patient is in emergent need of medical care. The most common way for this progression to occur is for a fatty plaque to rupture, leading to the formation of a clot that blocks blood flow. The first disease after this point is unstable angina (UA). As the restriction increases to a partial occlusion, the patient will experience chest pain that worsens without activity or that is not relieved with rest. Both stable and unstable angina present similarly in a patient. Typically, the patient will have episodes of chest pain that last from 3-10 minutes, but potentially lasting up to 30 minutes. This pain may radiate to the jaw, neck, shoulder, or arm. The patient will likely feel short of breath and may also experience nausea. If the patient takes a medication called nitroglycerine, the pain should resolve within 1-3 minutes. In UA, damage is still reversible, but intervention is emergently necessary to ensure that the disease does not progress. If UA progresses, it will progress to a condition commonly known as a heart attack, or in medical terminology as a myocardial infarction (MI). There are two types of MI: non-ST-elevation MI (NSTEMI), and ST-elevation MI (STEMI). In NSTEMI, some muscle in the heart has begun to die, and therefore at least some of the damage caused is irreversible. In a STEMI, there is a complete blockage of blood flow at some point and a large amount of muscle in the heart has begun to die. NSTEMI and STEMI are distinguished by characteristic findings on ECG; in a STEMI, the ST segment will be elevated above the baseline in some leads, while this elevation is absent in NSTEMI. The leads showing this change reflect the area of the heart impacted by the MI. This demonstrates the second need, longitudinal monitoring of continuous signals that can identify disease progression, and machine learning techniques that can account for the personal progression and varied rates of this progression. 

In order to prevent conditions such as stroke, interventions are necessary in known risk factor conditions, such as hypertension. Hypertension can lead to stroke in multiple ways. Very high blood pressure raises the risk of hemorrhagic stroke, as blood vessels in the brain may not be able to support higher pressures. Additionally, chronic hypertension is the main risk factor associated with ischemic stroke. The diagnosis of hypertension requires repeated blood pressure measurements (sustained hypertension), as measured by ambulatory blood pressure measurements.  Various reasons for blood pressure elevation must be identified, including white coat hypertension (when the blood pressure is elevated during a visit to a doctor but normal when measured in home settings), masked hypertension (when blood pressure is regularly elevated but detected as normal during a visit to a doctor), and evaluation in changes of blood pressure when sleeping versus when awake (nocturnal nondipping hypertension). Hypertension typically does not manifest with any symptoms, as the body is very good at masking the feeling of this pressure. Although high blood pressure has been colloquially associated with stress, headaches, or dizziness, these symptoms are typically not caused by chronic hypertension. The primary sign (and part of the diagnostic criteria) of hypertension is an elevated blood pressure. For diagnosis, at least two measurements on two different occasions of blood pressure above 120/80 mmHg are required. More recently, guidelines have suggested measuring blood pressure with an ambulatory blood pressure monitor over a 24-hour period, measuring blood pressure every 15 minutes during the day and every 30 minutes during sleep at night, and using the average values to have a better understanding of a patient's blood pressure \cite{RN19}. This sustained elevation may result in stiffer arteries, reducing arterial compliance. Additionally, over time, this chronic elevation may result in left ventricular hypertrophy seen on ECG or in changes in the retina. Most patients with hypertension are largely asymptomatic, with the chief clinical sign being that of elevated blood pressure.  When symptoms of hypertension do manifest, they are largely caused by organ damage that results from chronically elevated blood pressures.  Chronically elevated blood pressure can lead to heart damage, as the heart must work harder than normal to produce these elevated pressures.  This can lead to HF as the heart gains mass and loses efficiency, or to CAD as the increased mass of the heart requires increased myocardial oxygen supply.  Chronically elevated blood pressure can also lead to damage of the arteries.  This can lead to atherosclerosis, where plaque buildups can compromise coronary arteries, leading to CAD or cerebral arteries, leading to stroke.  Weakening of arterial walls can lead to kidney disease or to retinal disease.  Advanced hypertension can cause changes to the eye that can be observed visually by a physician. The definition of high blood pressure has undergone changes in recent years, with the SPRINT trial indicating that aggressive treatment of blood pressure to <120/<80 mmHg is associated with decreased mortality \cite{RN68}. The potential measurement of blood pressure from new sensing modalities can enable analytic techniques to identify cases of hypertension and evaluate the effectiveness of medication on reducing blood pressure, such as in the SPRINT trial. This illustrates the third need, where machine learning techniques, trained on continual data captured from new sensing modalities (the prior two needs), must provide actionable, interpretable estimations of signs, symptoms, and disease progress, in order to help guide treatment decision making and evaluate treatment effectiveness both prior to a diagnosis of a cardiovascular disorder and in the treatment and evaluation of recovery from an adverse cardiovascular event.

In the following sections, we explore the state of the art in technology associated with each of the needs. This survey reviews the technology available, the gaps that remain in addressing the needs, and highlights opportunities for researchers within the smart health field to design solutions with impact to clinical decision-making problems.

\section{New Sensors, Trends in Longitudinal Capture, Missing Data, and Sensor Selection}

The management of HF, CAD, and stroke can benefit from new sensing techniques by capturing acute data as well as detecting changes in sensed data over time. Each has unique signs and symptoms that manifest through a variety of changes in the body. For HF, improper blood flow can result in fluid retention (edema) in the lungs or the periphery, as well as develop signs of heart remodeling. Heart remodeling can be evidenced by third and fourth heart sounds (S3 and S4), as well as by a laterally or inferiorly displaced point of maximal impulse (PMI) of the heart on physical exam; the place where the heartbeat can be felt most strongly will migrate down and to the left of the thorax. One way in which improper blood flow can be detected is that the extremities will be cooler than normal. 

In CAD, stable and unstable angina will often result in physical pain felt by the patient in an episode that may last up to 30 minutes in the chest that may also radiate to the jaw neck and arm. The patient's heart rate and blood pressure will initially be elevated, although these can potentially decrease in NSTEMI and STEMI as the heart fails to operate optimally. The patient will breathe more quickly and will put more effort into breathing. Additionally, abnormal sounds may be heard with a stethoscope. It is possible for rales, an abnormal lung sound, to be heard at the posterior base of each lung. During chest pain, an ECG will show ST-segment depression, but this will change and progress to ST-segment elevation in STEMIs.

For stroke, this work focuses on the signs and symptoms that might lead to a stroke. Atrial fibrillation (AFib) is a relatively common arrhythmia that increases risk of stroke. AFib results when the atria of the heart beat ineffectively and randomly, causing turbulence within the atria. This turbulent flow allows for clots to form within the atria. If these clots are dislodged, they may travel through the arteries and become lodged in the brain, causing an ischemic stroke. AFib is classically defined as an ``irregularly irregular'' beat- the beat is not a typical rhythm (irregular) and additionally has no pattern determining when beats occur (irregularly). This is most often seen as absent P waves on ECG with variably occurring QRS complexes over a noisy baseline. However, this pattern could be detected by many techniques that measure pulse. Chief risk factors that predispose patients to AFib are age, other heart disease, diabetes, and chronic lung disease. HTN can also lead to stroke in multiple ways. Very high blood pressure raises the risk of hemorrhagic stroke, as blood vessels in the brain may not be able to support higher pressures. Chronic hypertension is the main risk factor associated with ischemic stroke.

These cardiac conditions present a range of sensing opportunities:
\begin{itemize}
    \item Acoustic measurement: capture of heart sounds to identify specific classes as well as respiratory effort are important in understanding acute conditions and changes in heart function over time. This also includes respiratory distress when lying down, causing patients diagnosed with HF to need to sleep in a more upright position.
    \item Electrical Measurement: Remote ECG measurements can identify periods of atrial fibrillation and other arrhythmias or help identify progression of CAD during an acute event.
    \item Blood flow: Understanding cardiac output, as well as measurement of blood pressure, is an important risk factor that needs periodic measurement.
    \item Fluid Retention/Weight Change: HF often results in lung and peripheral edema that results in swelling and can be measured by cooler temperatures in the periphery and changes in weight. 
    \item Physical activity and pain: In all cases, patient self-reported pain, fatigue, and physical activity may be surrogates for worsening conditions. Activity recognition can include posture detection to link with respiratory measurements.

\end{itemize}

\subsection{Existing Technologies and Applications}
\subsubsection{Acoustic Sensing/Vitals}
Vital sign monitoring has been explored through a variety of technologies. Each sensor type has been designed to address some of the sensing needs described in the previous section in an effort to replace or replicate tools available in acute care settings for remote environments. The stethoscope is one of the oldest such tools in medicine and is an implementation of acoustic sensing. By hearing and interpreting sounds from the patient, the physician can develop insights into the health of the patient and the functionality of the organs. Recently, digital stethoscopes have been utilized to better capture sounds. Digital stethoscopes provide benefit in allowing soft sounds to be more easily heard, but also allow for recording of sounds for later manual or computational analysis. As physicians have grown more reliant on advanced imaging techniques such as ultrasound, physical exam skill, including skill at auscultation, has decreased \cite{RN10}. 

Developing a digital stethoscope involves multiple components requiring heart sound capture, segmentation of the audio signal, and understanding of the cardiac cycle, best paired with an external signal such as ECG or pulse to determine the reference interval \cite{RN11}. A limitation here is that the time from electrical activity to sound production is not constant in all samples. Direct segmentation techniques involve utilizing Shannon energy to calculate an envelope and to find its peaks, and then use those peaks to reconstruct the cardiac cycle. Techniques applied to analyzing this data include machine learning techniques to classify the sounds, including support vector machines (SVM), artificial neural networks (ANN), hidden markov models (HMM), and gaussian mixture models (GMM), for identifying sounds and identifying next likely sound given the state in the heart beat cycle currently detected. These techniques have accuracies near 90\% for classifying signals as either normal or as having aortic or mitral valvular lesions. 

Work has also been done to develop low-cost devices that can act as a bridge between a traditional stethoscope and a cell phone \cite{RN64}. Constructing a cavity with good resonance is necessary in collecting good quality sound transmissions from the stethoscope. In particular, Sinharay et al. have evaluated using different kind of sensors to capture sounds to be transmitted from and to smartphones for analysis. 

In addition to detecting abnormal sounds in the cardiac cycle, there has been successful work in eliciting heart pathology from abnormalities within normal heart sounds. In particular, the normal cycle is composed of two sounds- one from the aortic valve closing and the other from the pulmonic valve closing. Both happen at nearly the same time, typically creating a single sound. However, some heart pathologies can impact the time between these. In a study of pediatric patients, high pressure in the pulmonary vasculature was found to be predicted by certain aortic and pulmonic valve relative intensities \cite{RN63}. Although this work has not been applied to adult patients, it could theoretically help to elicit information about the pressures at different points within the heart.

In several cases, radar has been utilized instead of direct, on-body measurement for detecting vital signs. Radar is able to detect periodic changes caused by both breathing and the heart, allowing heart rate and respiratory rate to be detected. Vinci et al. described a remote sensor that uses a six-point radar to monitor respiration and heartbeat \cite{RN60}.  It uses a continuous 24 GHz wave and a radiated power of less than 3 microwatts. It captures these values noninvasively in patients at rest. This is notable as it is a sensing modality that does not require attaching sensors to the human body.  This is particularly valuable in infants, in adults in severe conditions that cannot have additional attachments placed on the body, and as a modality that improves patient quality of life by limiting on-body sensors. The sensor designed in this paper does not have the limitations of other radar systems that require a wide frequency band to achieve more accurate results. Because of the six-point receiver architecture, this sensor can accurately measure angle and displacement by only measuring phase difference in backscatter patterns. Models regarding the permittivity of the skin allow them to estimate that their signal has 1.52 mm penetration as well as estimates of blanket and clothing impact. As a result, they can estimate where the edge of the torso is to aid in monitoring breathing. This provides an opportunity to noninvasively measure respiration and heart rate. However, it requires known, fixed postures of the individuals. Additionally, it will only work for one patient at a time. While this modality provides activity, displacement, and vitals monitoring in controlled, clinical environments or within specific remote environments (such as in the bedroom while asleep), it does not provide flexibility while moving. There are needs to extend such sensing systems to a variety of environments.

Work by Li et al. explore the use of radar technology for vital sign monitoring \cite{RN59}. Their system uses a hardware-controlled clutter cancellation system. This allows their radar technology to identify the difference between the person being monitored and background clutter that are likely present in rooms the person would be in. Authors propose taking ka-band radar systems that are meant for motion sensing and modify them for vitals sensing. Authors discuss existing work, design considerations for advancements, then opportunity to extend this to infant monitoring. The advancements in radar usage have come through the detection of the right frequency band to use. Different frequencies were shown to be able to go through different rubble with and without metal mesh. Authors then discuss the chip-level decisions that need to be made to create CMOS Doppler-based motion detectors. This allows vital sign detection through obstacles which can be important for noninvasive monitoring and for detection of vitals in emergency disaster scenarios. The application, however, is not clear for advanced signal processing of multiple vitals.

\subsubsection{Electrical Measurements}
Remote ECG monitoring has been explored by a number of researchers, primarily to solve the challenges that arise in noisy measurement. One issue that arises in continuous ECG monitoring, as with wearable ECG implementations, is that signals are often hidden by the noise of activity. Li et al. presented an approach for quantifying this noise \cite{RN54}. While earlier approaches focused on labeling ECGs as either clean or noisy, the approach presented by Li introduced five classifications, each with different amounts of information available to be extracted from the ECG. They defined the noisiest strips as those where artifact obscures signals to the point that there can be no confidence in any interpretation of the ECG. Strips with severe noise were those where some interpretation could be made, but interpretations could be confused as to where the QRS complexes fell or to whether ventricular flutter rhythms were present. In strips with moderate noise the QRS complex and presence or absence of ventricular flutter rhythms could be assessed, but finer signals such as P or T waves could not be extracted. Minor noise was the label given to strips with some amount of noise, but where P waves and T waves could be extracted. This level of noise allows for the analysis of atrial arrhythmias such as atrial flutter. Finally, clean ECGs were those where no noise was present. The authors produced training data by adding three types of noise to the original clean dataset: baseline wandering, electrode motion, and muscle artifact. They trained an SVM to classify strips based on the amount of noise present and validated this classification scheme on real noisy data. This validation showed good agreement between manually annotated labels and model output labels, with the greatest confusion present where samples had been manually annotated as having minor noise, but the model labeled the samples as having moderate noise. The authors note that a chief limitation of this work was that the model was not trained for or with an arrhythmia database, which substantially lowers its effectiveness on samples with arrhythmias. Additionally, they note that methods based on continuous features rather than discretely extracted features would be likely to show greater performance.

Once identified, several approaches have been implemented in order to account for and to correct motion artifacts. Sriram et al. addressed this problem by utilizing a triaxial accelerometer \cite{RN53}. ECG signals are usable as a means of continuous biometric security. However, this continuous security is lost when the ECG signal is distorted with motion artifact. This approach shows that supplementing the raw ECG signal with features extracted from acceleration allows for accurate classification of ECG subject identity. They segmented signals to windows containing roughly four heartbeats, averaged those four beats together, and then corrected for baseline abnormalities with linear interpolation of q-minima and a high pass filter in association with the accelerometer features. These features then served to correctly identify users using either a k-nearest neighbors or a Bayesian network classifier.

Another issue that arises with automatic ECG monitoring is that many abnormalities might be troubling in one patient while normal in another. Chen et al. \cite{RN52} described an approach to train ECG monitoring systems to discover patient-specific abnormalities. This work utilized an accelerometer to reduce the number of false alarms in monitoring systems. Over time, this system learns the normal for a given patient and uses a knowledge of this normal in order to reduce false alarms.

\subsubsection{Blood Flow}
The American College of Cardiology and the American Heart Association (ACC/AHA) recently released guidelines that suggest ambulatory blood pressure measurements, those taken at home in 15 minute intervals including during sleep, should be captured to better understand a patient's blood pressure and potential cardiovascular risks \cite{RN87}. The sphygmomanometric and oscillometric techniques are well-established as the predominant means by which blood pressure is typically measured \cite{RN51}. Both methods involve the inflation of a pressurized cuff, typically around the patient's upper arm and maintained at the level of the heart. The pressure in the cuff is increased to above realistic values of the systolic blood pressure, and then slowly decreased. In the auditory sphygomomanometric method, sounds called Korotkoff sounds can be heard just distal to the cuff as it deflates. The pressure at which these sounds are first heard is the systolic pressure, and the pressure at which these sounds are no longer heard is the diastolic pressure. In the oscillometric technique, minute variations in pressure as the heart beats against the pressurized cuff are measured and the systolic and diastolic blood pressures are extracted from these variations \cite{RN50}. Most at-home blood pressure monitoring devices utilize the oscillometric technique, which is well-validated to have performance similar in quality to the sphygmomanometric technique \cite{RN49}. Recently, cuff-less blood pressure monitoring techniques have been explored in order to record blood pressure.

The most common cuff-less approach thus far is to use photoplethysmography (PPG) and ECG to capture pulse arrival time, pulse transit time (and pulse wave velocity), as surrogates for blood pressure, then use analytic techniques to estimate the systolic and diastolic blood pressure values \cite{RN88, RN89}. If the posture of an individual is known, these techniques are able to measure an estimate of the blood pressure, without disturbing the individual with frequent cuff inflations. However, the ECG and PPG combination can result in error in blood pressure estimation because it does not appropriately account for artifacts that exist between the ECG measurement of a pulse and the PPG capture of the pulse arrival time \cite{RN90}. To account for this, researchers have turned towards dual PPG capture \cite{RN91, RN92} over a small portion of the artery to account for pulse transit time, or bio-impedence measurements, which are better able to locate the artery and avoid capturing blood profusion time into capillaries \cite{RN93, RN94}. Ballistocardiogram approaches look to capture pulse arrival time through the small changes in pressure sensed by the waves in each pulse, providing a method for capturing cuff-less blood pressure whenever participants are still \cite{RN95, RN96, RN97}. These approaches all look to address cuff-less blood pressure when the participant is in a fixed, known position, and provide the opportunity for more frequent ambulatory blood pressure measurement.

\subsubsection{Fluid Retention}
\label{sec:fluid}
While prior studies, such as Tele-HF and Beat-HF, attempted to use weight scales as a surrogate for fluid retention in HF, the measurement of 3 pounds of weight change was not an alert that was able to reduce HF readmissions \cite{RN71, RN34, RN33}. A number of attempts to measure peripheral edema and fluid retention have focused on the development of smart socks that look to measure fluid buildup in the ankles \cite{RN41, RN98}. A stretch sensor measures the expanding duration of the patient's ankle both as edema increases throughout the day and as edema increases over time. The context-awareness allows the device to discard ankle measurements when motion, muscle contractions, or an incorrect posture would interfere with the measurement. This sock was able to reliably determine the participant's posture, and measurements of fluid retention were well correlated, but additional study is needed to determine if this measurement is accurate enough, and whether it can generate alerts early enough to intervene in HF patients. Yao et al. came to similar conclusions of needing further study of their sensor to classify edema \cite{RN99}, as this remains an open area of research.

\subsubsection{Physical Activity and Posture}
Activity, posture, and pain are important measurements in understanding symptom and treatment effectiveness in patients diagnosed with cardiovascular disorders. Measurement of respiratory distress in HF patients requires a measurement of posture, measurement of blood pressure through proxy measures such as pulse transit time require a measurement of posture, as did the smart sock for fluid retention (Section \ref{sec:fluid}). While each sensor can capture posture, smartphones excel at this \cite{RN100}, often coupled with other applications tracking activities of daily living \cite{RN101, RN102}. Recently, smartwatches have shown to accurately detect postures and exercises \cite{RN103, RN104}, which is important for patient monitoring, since smartphones are often in the proximity of the user, but often not physically on the user, unlike smartwatches \cite{RN105}. These can also provide important context to the measurements captured by the other modalities discussed in this section \cite{RN37}.

\subsection{Gaps}
As richness of data increases across the variety of sensors, the potential for noise and missingness increases as well. It is difficult to understand the context in which measurements are captured. Accuracy of posture detection and presence of other noisy attributes impact the potential success of different sensing modalities. It is also unlikely a patient will wear all sensors all the time, as this will provide excessive burden. While a measurement performed on occasion is likely to be a high-quality measurement, continuous and automated measurements introduce a greater deal of variability in the quality of measurements. For instance, a once-a-day measurement is likely to be a measurement where the patient will intentionally position themselves appropriately and remain motionless during the measurement. A patient monitoring their blood pressure will likely sit upright with their legs uncrossed, or a technician performing an ECG will ensure that the printed ECG is taken at a point where the patient is motionless, and no artifacts are present. Conversely, more frequent or continuous monitoring must account for noise introduced by motion artifacts as well as from noise introduced from other suboptimal measuring conditions. As such, a number of challenges remain in capturing the necessary signals:
\begin{itemize}
    \item Acoustic measurement: challenges in identifying the specific patient limit the potential for non-wearable sensors. Wearable sensors must account for noise across a variety of motions, environments, and potential sensor misplacement.
    \item Electrical Measurement: Continuous ECG requires multiple leads to be worn at the same time. Devices such as the Apple watch provide potential for requesting ECG periodically when other sensing modalities dictate when it is necessary \cite{RN85}, but the correlation between these modalities and necessary ECG readings has not been well studied outside of AFib.
    \item Blood flow: Cuff-less blood pressure monitoring must extend to continuous, beat-to-beat measurements without constantly restraining users to fixed, known postures.
    \item Fluid Retention/Weight Change: Edema measurements have not been clinically validated to show the degree of fluid retention which must generate alerts that can clinically improve outcomes.
    \item Physical activity and pain: Remote measurement of acute and chronic pain remains an open challenge.
\end{itemize}{}

\subsection{Opportunities}
An additional source of noise can be introduced by the redundancy of signals that can exist. Different physical phenomena can be measured by different modalities, many of which will produce slightly different readings. Heart rate can be derived from multiple sources: auditorily by stethoscope, electrically by ECG, visually by PPG, and visually by radar. It stands to reason that these redundant values could be exchanged for each other, but that exchange may not completely be a one-to-one relationship. Transfer learning is an ongoing field of study that seeks to apply existing models to data that was not used in training or was only used minimally in training \cite{RN106, RN107, RN108}. Transfer learning could be applied to this problem as a way to apply a single model to patients with disparate data collection modalities.

Missingness in data also increases as richness increases. While binary parameters used in many risk models (e.g. history of HF, current diabetic status, etc.) are easy to collect and even possible to impute, continuous monitoring opens the possibility of more complicated missingness. A battery may fail on a sensor leading to a variable period of missingness. Wearable sensors may introduce missingness secondary to poor compliance or poor utilization. The missingness introduced by gaps in continuous monitoring is more difficult to impute and presents a challenge in building comprehensive models \cite{RN109, RN110}. A number of opportunities emerge for immediate and impactful research on sensing signs and symptoms of cardiovascular disorders, illustrated in Figure \ref{fig:sensors}, and listed below:

\begin{figure}[ht!]
  \centering
  \includegraphics[width=\linewidth]{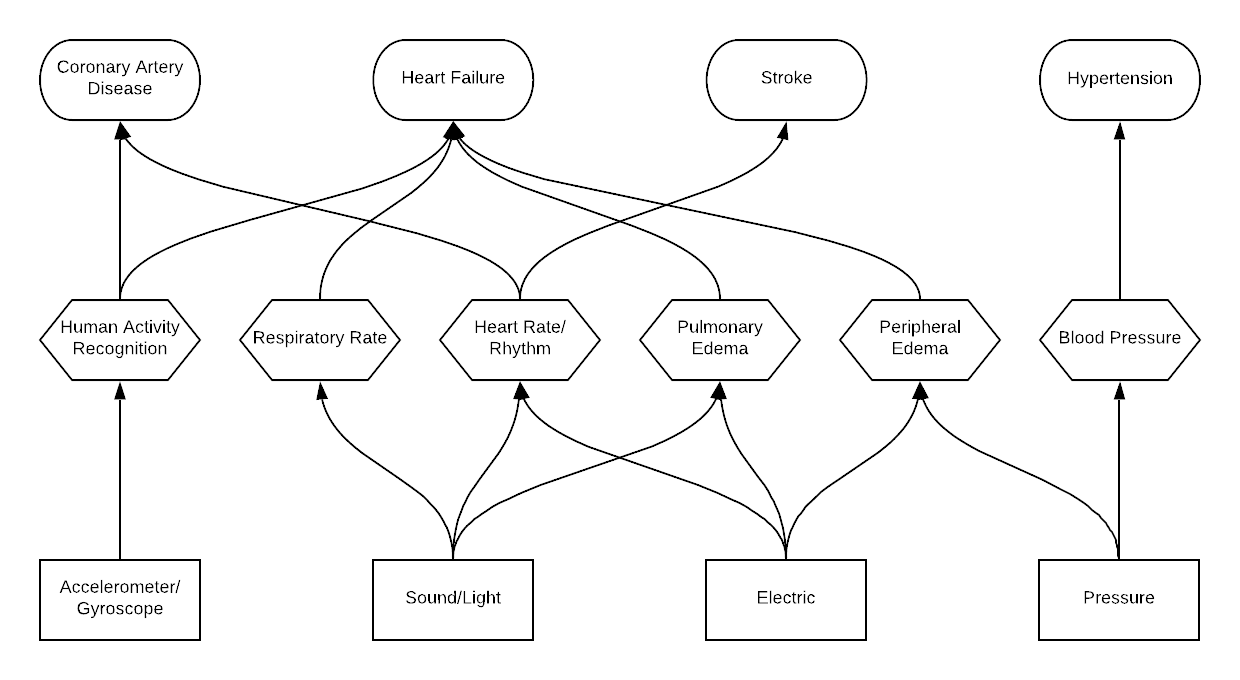}
  \caption{Overview of basic sensor categories proceeding to physiologic value measured and to overarching disease state.}
  \Description{Overview of basic sensor categories proceeding to physiologic value measured and to overarching disease state.}
  \label{fig:sensors}
\end{figure}

\begin{itemize}
    \item Integration of multiple sensing modalities into a single platform, reducing the number sensors needed to be worn. High impact areas appear to be the wrist (smartwatch) and chest (heart and lung sounds).
    \item Integration of machine learning techniques to help identify when longitudinal data capture is necessary, similarly to ECG requests to verify periods of arrhythmias associated with AFib detection with the Apple Watch \cite{RN85}.
\end{itemize}{}

\section{Continuous Data Collection and Analytic Models}

Beyond the acute sensing and detection of symptoms related to HF, CAD, and stroke, analytic opportunities arise in the processing of this data longitudinally and continuously. As discussed, the progression of CAD from stable and unstable angina to NSTEMI and STEMI represent longitudinal changes that may have periods of rapid change interspersed. Similarly, untreated hypertension can lead to stroke if untreated. Changes in heart remolding in HF may be represented by changes in heart sounds as captured by acoustic sensing.  Patients living with HF may experience long term changes in the amount of physical exertion required to perform activities of daily living.  These changes may be gradual and unnoticeable to the patient, but may represent worsening condition or recovery. 

These cardiac conditions present a range of analytic techniques necessary to capture longitudinal changes in continuously-sensed data:
\begin{itemize}
    \item Continuous capture of acoustic sensing: Understanding how sounds change over time may allow for the identification of new signals that represent earlier identifiers of worsening conditions or treatment effectiveness.
    \item Continuous capture of electrical signals: While the detection of arrhythmias may be present in surrogate measures such as heart rate, detection of changes in ST segments of an ECG may allow for early alerts and acute care.
    \item Continuous capture of vitals signals: Understanding the changes in the variety of vitals signals captured and how they may relate to each other can provide an understanding of improving or worsening risk factors relevant to HF, CAD, and stroke.
    \item Continuous capture of physical activity: Physical activity and sleep are important functional measures of recovery, and accurate, longitudinal understanding of functional change can be correlated with improved mortality and prevention of adverse events.
\end{itemize}

\subsection{Existing Technologies and Applications}
\subsubsection{Continuous Capture of Acoustic Sensing}
Electronic auscultation is useful for deriving characteristics of other parts of the cardiovascular system than sounds generated specifically by the heart valves. A carotid bruit is a sound created by turbulent blood in a carotid artery, often caused by narrowing that in turn is produced by atherosclerotic plaques. Knapp et al. looked at the effectiveness of carotid bruit detection by electronic auscultation \cite{RN62}. Out of 1,371 patients in this study, 84 were found to have carotid bruits by electronic auscultation. These patients were matched with controls who did not have bruits, and both patients from each pair were assessed with duplex ultrasound to determine extent of carotid stenosis. Bruit detection with electronic auscultation and manual annotation was found to have a sensitivity of 88\% for stenosis $\ge$ 50\%, and a specificity of 58\% with duplex ultrasound providing the ground truth.

Work by Palaniappan et al. surveyed machine learning techniques to further analyze lung sounds \cite{RN61}. They evaluated 59 papers that used signal processing and machine learning techniques on a variety of lung sound problems including normal breath sounds, abnormal breath sounds, and a series of sounds called adventitious lung sounds. This survey highlights an important need by evaluating short term sounds, long term sounds, and identifying normal and abnormal sounds across the different time periods. Most works in this survey focused on specific frequencies (between 150 and 2000 Hz, though they found that most work typically worked at 150 Hz), and evaluated machine learning techniques such as k-nearest neighbor, ANNs, HMMs, GMMs, genetic algorithms, SVMs, and fuzzy logic to classifying a variety of lung sounds. They found that by using piezoelectric microphones, contact microphones, and electric microphones, and one commercially available lung sound instrument, they could design electronic stethoscopes that filtered out heart sounds to capture necessary lung sounds. Similarly, one could use the same techniques to filter out the lung sound to capture the heart sounds. Using standard time-domain and frequency-domain signal processing features, algorithms were able to classify lung sounds with between 83-93\% accuracy. Deep neural network techniques, such as convolutional neural networks (CNN), recurrent neural networks (RNN), and long short-term memory networks (LSTM) should be able to extract more time and frequency domain features rather than hand crafted features explored by authors to achieve higher accuracy.

\subsubsection{Continuous Capture of Electrical Signals}
In clinical settings, most ECGs are performed as 12-lead ECGs. In these ECGs, there are 10 electrodes attached to the patient and 12 different measurements taken from these electrodes. Each provide a one-dimensional view of the magnitude of the vectors of all electrical impulses in the heart relative to a given axis. Different axes allow for information to be obtained about the functionality about different parts of the heart. Depending on the goals of remote monitoring, remote ECGs will typically only include a subset of these typical views. As a result, methods that can accurately detect essential signals from minimal lead ECGs are necessary.

Work by Jambukia et al. surveyed machine learning techniques to analyze and classify ECG signals \cite{RN58}. They evaluated 31 papers that used signal processing and machine learning techniques in order to extract clinically significant features from raw ECG signals. Most of the papers evaluated used the MIT-BIH arrhythmia dataset \cite{RN57} for both training and testing purposes. Two aspects of ECG classification considered were ECG beat classification for individual, isolated beats, and ECG signal classification for interpretation of a longer signal. Some approaches evaluated involve signal feature extraction followed by threshold-based algorithms such as the Pan-Tompkins algorithm. Other approaches utilized various neural network architectures, with the authors finding that of the architectures studied, multilayer perceptron neural networks provided the best performance. Recurrent neural networks, such as the long short-term memory (LSTM) architecture, were not evaluated in this survey. Deep learning techniques have also been utilized for ECG evaluation. Yildirim showed that a bidirectional LSTM architecture can reliably classify five different rhythms from the MIT-BHI arrhythmia database \cite{RN56}. This bidirectional LSTM model achieved accuracies greater than other techniques. Additional deep learning techniques that combine CNN and LSTM have been used to detect AFib without explicit feature extraction (such as R peak extraction) \cite{RN111}. Further deep learning techniques have looked at a variety of processing individual beat anomalies and sequence anomalies \cite{RN112}, though time series presented to CNN models often needs fixed windows of time to be pre-determined for evaluation. Additionally, some work uses a single lead \cite{RN113} for detecting arrhythmia, though it is likely at least two leads are currently necessary for other ECG feature extraction.

There is evidence to suggest that patients at risk of cardiac pathology benefit from more continuous remote ECG monitoring. The mHealth Screening to Prevent Strokes (mSToPS) randomized clinical trial is an ongoing trial of 2659 patients investigating the benefit of continuous monitoring for AFib \cite{RN55}. As reported by Steinhubl et al., the initial phase of the trial discovered that for individuals at risk of AFib, home ECG monitoring was superior to routine care for discovering new incidence of AFib. In the actively monitored group, there was a 3.9\% diagnosis of new-onset AFib, vs 0.9\% in the control. This resulted in earlier initiation of anticoagulative therapy (a preventative measure for stroke) in these patients. However, this has also resulted in a higher healthcare utilization among these actively monitored patients. This trial is still ongoing- the ultimate clinical impact is still unknown. Clinical outcomes are due to be published in a 3-year follow-up.

\subsubsection{Continuous Capture of Vitals Sensing}
Ultasonography is a technique that uses ultrasonic sound waves to produce images of tissues beneath the skin. Ultrasonography is valuable for visualizing structures that are unreachable noninvasively. In hospital settings, point-of-care ultrasound has increasingly grown in utilization as mobile ultrasound systems become cheaper and comparable in quality to larger ultrasound systems \cite{RN48}. Point-of-care ultrasonography is useful as a tool that physicians can bring to the bedside for aid in diagnosis, much like a stethoscope, but deep learning techniques are necessary to evaluate the ultrasound images and classify changes in conditions.

Ultrasonography can also be used to evaluate the fluid status of the lungs. As described in Assaad et al., lung ultrasound is a valuable tool for quickly assessing the health of a patient's lungs \cite{RN46}. Certain visual findings, such as ``B-lines'' are highly associated with edema and various pulmonary pathologies. These visual findings also change very rapidly, reflecting the present disease state more accurately in some cases than measures such as blood oxygen saturation. Lung ultrasonography is also useful in differentiating between cardiogenic and noncardiogenic pulmonary edema; cardiogenic pulmonary edema typically shows more uniform findings and plural effusion (fluid buildup in the tissue surrounding the lungs). Lung ultrasonography is an underutilized technique in medicine and lacks standardization in training and implementation.

Work by Bhuyan et al. explores an exciting possibility of wearable ultrasound for the monitoring of internal function noninvasively \cite{RN45}. In order to create a small form factor that could be used to measure organ function with wearable, remote ultrasound, they created a small, flexible probe through a flexible PCB integrated circuit. They also used a system that has only one transmit and one receive channel to avoid excess signal degredation. This system has a bandwidth of 10 MHz, power consumption of 6.72 mW per channel, and uses 16 such channels to measure a 5.6 mm x 1.6 mm area. They used classical image processing with ultrasound for their validation. Their system, however, used an attached cable to measure. There is an opportunity to create a remote, continuous version of such a system if a flexible PCB-based wearable ultrasound with necessary battery and wireless transmission capabilities were added, but, computer vision techniques are needed to enhance the analytic component of the wearable ultrasound.

Echocardiography is the practice of using ultrasound in order to visualize the structures of the heart. Echocardiography can take place either as an invasive transesophageal echocardiography (TEE) or as the noninvasive transthoracic echocardiography (TTE). Many different aspects of the heart can be described and quantified via echocardiography \cite{RN47}, including size, function, and mass of various structures of the heart. Measurement of these parameters aids in the diagnosis of HF. For instance, left ventricular mass or poor emptying are markers of HF. Valvular disfunction, such as stenoses or regurgitations can be directly observed. These measurements also aid in assessing cardiac function in CAD, particularly following MI; injured portions of the cardiac wall will often move less than they normally would.

Various groups have found preliminary success in applying deep learning computer vision techniques to the analysis of ultrasonographic images. The first step in automatic analysis of ultrasonographic images is to recognize the view in question. {\O}stvik et al. describe the use of a convolutional neural network (CNN) to classify TTE images according to the view being presented \cite{RN44}. This method showed classification high accuracy in distinguishing among seven different TTE views. Additionally, the authors described a technique for extracting 2D slices from 3D images and achieved a mean error of 4\textdegree.

Techniques for measuring edema include cuffs that track ankle circumference and measurement of electrical impedance. Weight monitoring is sometimes used as a proxy for tracking edema, as edema co-presents with fluid retention. There has been success in implantable impedance monitors to measure pulmonary edema. Yu et al. found that intrathoracic impedance serves as a predictor for imminent hospitalization due to fluid overload \cite{RN43}. In a population of 33 patients with HF, a device consisting of a pacemaker and defibrillator was implanted. The device measured the impedance between those two leads. This study found that there was a significant decrease in impedance prior to hospitalization with fluid overload. This decrease began on average two weeks prior to hospitalization and continued through the date of hospitalization. 

Impedance monitoring has also been implemented in noninvasive and ambulatory monitoring systems. Weyer et al. describe a system that incorporates both ECG and noninvasive impedance cardiography \cite{RN42}. This device includes Bluetooth connectivity and a battery that lasts for up to 21 hours. This system could be implemented for long-term monitoring in patients with HF to monitor pulmonary congestion and to potentially allow remote interventions before hospitalization is necessary.

The internal and external jugular veins provide drainage from the head into the heart. The right jugular veins are positioned almost directly above the right atrium, and therefore the pressure within them is very closely tied to the pressure of the right atrium. The external jugular vein's filling level indicates the pressure within the right atrium and will be distended in cases of right heart failure. Pulsations can be observed with great difficulty in the internal jugular vein. These pulsations provide evidence as to the relative timing and forces involved in right atrial contraction, atrial relaxation, right ventricular contraction, venous filling after the closing of the tricuspid valve, and emptying of the atrium after opening of the tricuspid valve.

As venous pressures are so much lower than arterial pressures, measurement of the jugular venous pulse is much more difficult than the measurement of arterial pulses. However, Amelard et al. were recently able to utilize a technique called PPG Imaging (PPGI) as a viable technique to correctly measure the jugular venous pressure \cite{RN35}. This technique uses a system located approximately 1.5 meters away from a supine patient. A light shines on the patient and the reflected light is analyzed to identify pulsations. The arterial pulsation from the carotid artery is easier to detect, and the jugular pulsation can be identified as a corresponding inverted pulsation at a location near but lateral to the arterial pulsation. In this study, the ground truth arterial waveform was verified with a PPG measuring device. Pertinent clinical features were consistently able to be extracted from the venous waveform, including the c, v, x, and y waves (corresponding to the contraction of the right ventricle, systolic filling of the right atrium, relaxation of the right atrium, and beginning of the filling of the right ventricle). In about half of subjects, the a wave was also observed (corresponding to the contraction of the right atrium). The ability to regularly monitor and quantify these waveforms could allow for new techniques in monitoring right heart function. Ballistocardigrams may be another way to measure this venous pressure.

Signals that capture continuous blood pressure, described in Section 3, may also be extended to capture a variety of heart rate, heart rate variability, blood pressure, respiratory rate, and changes in these values. Obstructive sleep apnea, a condition in which airwaves are restricted causing the body to wake up from sleep to begin breathing again, increases heart rate, respiratory rate, and blood pressure, keeping patients from falling asleep. This has a direct relationship with blood pressure and nocturnal nondipping hypertension, and treatments for apnea have shown to be correlated to improvements in blood pressure \cite{RN114}. This means approaches for measuring cuff-less blood pressure cannot be restricted to periodic, ambulatory measurements, but must transition to continuous beat-to-beat measurement and interpretation.

Finally, telemonitoring trials for HF readmission have tracked longitudinal measurements of symptoms, vitals, and patient qualitative reports \cite{RN71, RN34, RN33}. In the Tele-HF study, a number of vitals signals and patient reported outcomes generated alerts for interventions if the values were below a specified threshold, or represented a significance drop from the prior day's values. However, the study was unable to find a statistically significant reduction in readmissions in the intervention arm. Ong et al. in the Beat-HF study tried to use some machine learning techniques to further identify risks of adverse events, and while the study was unable to reduce readmissions, the techniques did show some promise in stratifying patients \cite{RN115}, as did further statistical techniques applied to the Tele-HF data \cite{RN34, RN81}. With the addition of more signals captured, and techniques that can better account for varied time-domain aspects of analytics, it is possible that better just-in-time alerts can be generated for preventing future recurrent HF events.

\subsubsection{Continuous Capture of Physical Activity}
For cardiovascular disorders, the detection of activities and postures is important in understanding the other biomarkers captured, providing context for their readings. For example, at nighttime, knowing the posture of the user provides context for dyspnea measurements and heart sound recordings for HF patients. In addition to providing context for the other vitals measurements related to cardiovascular disorders, the change in physical activity performance can show increasing effects of HF symptoms, pain as a result of CAD, and acts as a surrogate for the general well-being of these patients.

Many research-oriented activity recognition platforms focus on the detection of activities of daily living \cite{RN37, RN40, RN39, RN38} and understanding daily exercise intensities. These sensors are capable of tracking sports movements in the healthy and measure sedentary time in the elderly, and come in many forms of smartwatches, smartphones, smartwatch-sized sensors \cite{RN103, RN40}, embedded within shoes, and most recently within eTextiles \cite{RN116, RN117}.

HF participants have had improved outcomes in mortality and readmission when involved in cardiac rehabilitation programs that encourage continuous physical exercise \cite{RN118}. This physical exercise routine has shown that measurements in improved peak exercise capacity correlate with improved cardiovascular outcomes. Home-based cardiac rehabilitation systems centered on physical activity detection in order to quantify a home-based exercise routine \cite{RN119}. However, such systems do not yet quantify improvement directly from physical activity measurements. This is necessary since adherence in cardiac rehabilitation programs is often quite low \cite{RN120}.

\subsection{Gaps}
Deep learning techniques have made the exploration of time series data more fruitful with the development of automatic features that represent longitudinal risk or outcomes. Techniques such as attention-based LSTMs have shown promise in exploring continuous time-series data to predict clinical mortality, decompensation, and length of stay, which outperform hand-crafted feature extraction, and other deep-learning techniques that do not find focus on specific periods of time, in intensive care unit data \cite{RN121}. However, these techniques have not been applied to this remote data yet because the integration of these sensing techniques have not yet occurred.

A number of the analytic techniques tied to the use of these new sensing paradigms have focused on the diagnosis of specific anomalies or classification of specific types of sounds or signals captured. What is still needed is the following:
\begin{itemize}
    \item Integrated sensors that can capture signals frequently, or continuously, over entire study periods.
    \item Machine learning techniques that can explore multiple windows of time over multiple combinations of available signals in order to quantify trajectories in signals, identifying longitudinal patterns and changes in signal that may be indicative or worsening conditions or treatment effectiveness and recovery, extending beyond anomaly detection and signal classification.
\end{itemize}

\subsection{Opportunities}
One way in which advanced work in analytics could be incorporated is in better personalized monitoring for risk of CAD. As previously discussed, CAD is a condition often characterized by gradual worsening of chest pain that culminates in a heart attack. Ideally, monitoring systems would be able to follow gradual changes prior to the rupture of a plaque that causes a heart attack. In the early stages of CAD, activity monitoring could be used to assess wellness. By tracking a certain threshold of activity that the patient does not (or cannot) exceed, a monitoring system can estimate the severity of angina. As that threshold begins to decrease, the patient's angina is likely increasing and greater intervention may be warranted. The monitoring system for a patient at risk of CAD should also include an ECG system to watch for changes associated with a heart attack. If any electrical changes concerning for a heart attack begin to appear in the patient's ECG, then emergency services would be required. Earlier interventions are associated with better outcomes, and a monitoring system like this coupled with improved analytics could potentially allow for earlier treatment, leading to less overall damage and better patient outcomes.

Improved analytics could also be implemented to better treat valvular diseases. Unlike the other pathologies discussed here, there are few risk models for predicting future valvular heart disease. However, advanced analytics could be implemented to allow for earlier detection of valvular disease. As discussed above, these abnormalities change the way in which blood flows through the chambers of the heart, producing turbulence that can be detected as sound. The most straightforward evaluation for valvular disease in remote wearable settings would involve electronic stethoscopes continuously monitoring the patient's heart sounds. By learning the normal sound profile of a patient, new changes and murmurs could be quickly identified. After identifying a particular valvular disease and its associated murmur, long-term monitoring with electronic stethoscopes could be used to characterize the severity of the valvular insult; most murmurs initially increase in intensity, but in later disease stages decrease in intensity. Rather than risking false negative screening in physical examinations, a longitudinal monitoring system could detect the changes along this trajectory to allow for more informed decision making. As a more advanced option in monitoring valvular disease, miniaturized ultrasound probes could be incorporated into a wearable system. These could be used for imaging and analyzing the valvular parameters such as cross section and flow.  Additional work into computer vision interpretation of ultrasound images would be necessary in order to automatically process these signals. Vital monitoring can also directly feed into an understanding of valvular disease. In particular, blood pressure can reflect aortic valve lesions. Finally, systems to monitor valvular disease could monitor symptomatic disease progression. As many types of valvular disease may ultimately lead to HF, the opportunities presented above for HF apply here as well. Foremost among them would be activity recognition, where late stage valvular disease can manifest with a loss of stamina in day-to-day activities.

Opportunities:
\begin{itemize}
    \item Improved Machine Learning Processing of Existing Sensor Modalities: Development of machine learning techniques that can extract meaningful data from non numerical sources, expanding on the computer vision work done in automatically processing and interpreting ultrasound images.
    \item Time-Series Machine Learning Models: Development of machine learning techniques that can process longitudinal data and account for multiple channels of data, sampled at different frequencies, and with different segment lengths of importance, are required to develop new risk prediction techniques and alerts based upon continuously captured data.
\end{itemize}

\section{Clinical Interpretability, Analytic Models, and Treatment Paradigms}

Clinical risk prediction models and those that predict adverse events have helped guide medical treatments and improve patient care. These techniques, with machine learning modeling, have the potential to improve clinical care in both the acute care settings \cite{RN122} and remote care settings. This includes understanding the diagnosis and progression of diseases and the personalized patterns and signals that can be captured by advances made in categories listed in Sections 3 and 4. Some preliminary work has been conducted in clinical trials on HF patients, understanding distinct patient phenotypes within the disease. In one such HF trial, clinically-distinct clusters of patients were found to have different time-to-event predictions and outcome rates \cite{RN123}. Another relevant clinical trial in HF patients is the Treatment of Preserved Cardiac Function Heart Failure with an Aldosterone Antagonist (TOPCAT) trial \cite{RN125, pitt2014spironolactone}. The purpose of this trial was to determine if a treatment designed specifically for HF patients with preserved ejection fraction could improve outcomes, a patient population where such treatments have not been found to universally treat these patients. This trial was also unsuccessful in showing that HF patients treated with Spironolactone had better outcomes \cite{pitt2014spironolactone}. However, due to some issues with data gathered in certain regions, investigators began taking a closer look at subsets of patients, to determine if specific patients were actually helped by the treatment. The investigators found regional variations lead to different treatment effectiveness in cohorts of participants \cite{RN124, RN125}. This indicates that HF patients diagnosed with preserved ejection fraction may benefit from cluster analysis, looking at personalized differences in outcome rates where different treatments may be helpful for different subsets of patients. These provide for the basis of the following needs in HF, CAD, and stroke:
\begin{itemize}
    \item Personalized models: as illustrated by the TOPCAT findings, these diseases are quite complex and understanding the person-to-person variation allows for specific risk prediction based upon data collected, along with matching techniques that allow for comparison to patients most similar to the individual modeled.
    \item Dynamic adaptation: models must account for the varied data types potentially collected, the varied rate at which they are collected, how well to link them to data gathered in acute care settings, and be able to update as a disease progression worsens or treatment regimen proves effective, including providing confidence metrics that suggest the collection of additional data, if necessary.
    \item Interpretable machine learning: as the data size progresses, medical models must be able to explain the driving risk factors in a manner interpretable to clinicians in order to guide treatment decision making.
\end{itemize}

\subsection{Existing Technologies and Applications}
\subsubsection{Risk Prediction Models}
Much of stroke risk prediction is tied to the risk associated with AFib. In particular, it may be appropriate for patients with AFib to undergo anticoagulation therapy in order to reduce their risk of stroke. Anticoagulation therapy is any therapy that works to reduce the rate at which blood clots form. This type of therapy can be beneficial by preventing thromboembolic stroke. Conversely, this type of therapy can be detrimental by promoting life-threatening bleeds, such as in hemorrhagic stroke. Therefore, implementation of any anticoagulation therapy must be implemented with great care. In addition to models that predict only stroke risk, the ACC/AHA Pooled Cohort Equations treat stroke and CAD together. 

CHA2DS2-VASc is a model that predicts 12-month thromboembolic event rate (including stroke, pulmonary embolism, and peripheral thromboembolism) in patients with AFib who are not undergoing anticoagulation therapy \cite{RN15}. Creation of this model drew upon the efforts of and improved upon multiple older models in order to apply more broadly and accurately to diverse patient populations. One chief exclusion in this model is that only patients with non-valvular AFib are considered. The parameters considered in this model are presence of HF, hypertension, age, diabetic status, history of stroke or other thromboembolic event, history of any vascular disease, and gender. This model aids clinicians in prescribing anticoagulants, which increase the risk of bleeds but decrease the risk of thromboembolic events (including stroke).

The HAS-BLED model was created to predict the risk of bleeding in anticoagulated patients with AFib \cite{RN126}. The parameters included in this model are hypertension, history of liver or kidney dysfunction, history of stroke, history of bleeding, difficulty calibrating oral anticoagulation therapy, use of alcohol, and use of certain drugs that may increase bleeding risk. Recommendations by groups such as the European Society of Cardiology \cite{RN16} are that CHA2DS2-VASc and HAS-BLED be used in conjunction for informed decision making, and that HAS-BLED alone should not be a reason to withhold anticoagulant therapy.

Other models have been produced to predict general risk of stroke. The MyRisk\_Stroke Calculator is a model to predict 10-year risk of stroke \cite{RN17}. This estimator was built on a prospective dataset where collection began in 1992 and validated with a second dataset with collection beginning in 1998. Follow-up was through the year 2007. In this cohort, the parameters found with an association to stroke risk were age, gender, education status, high blood pressure, smoking status, alcohol consumption, activity levels, anger, depression, and anxiety. Additionally, comorbidities such as renal disease, diabetic status, HF status, CAD, peripheral arterial disease were included as features in this model. The model was created as a Cox proportional-hazards model and predicts 10-year risk of any type of stroke.

Another stroke risk model is the QStroke score \cite{RN18}. QStroke was developed to be used for all patients without history of stroke but intended specifically to be used as a supplement or replacement for CHA2D2-VASc in predicting risk associated with AFib. The QStroke model features the following as parameters: age, gender, ethnicity, Townsend deprivation index (an index related to socioeconomic status), smoking status, body mass index, systolic blood pressure, blood lipid levels, and family history of CAD. Hypertension, diabetic status, AFib status, HF status, CAD, presence of rheumatoid arthritis, renal disease, and valvular disease were also included as pertinent comorbidities. The QStroke model was created as a Cox proportional-hazards model and predicts 10-year risk of any type of stroke.

Many attempts have been made to assess the risks of developing CAD. Among the most current of these are from the ACC/AHA Task Force on Practice Guideline \cite{RN19}. That work introduced a set of models termed the Pooled Cohort Equation to predict a primary CAD event within 10 years. The predicted risk in this model is based on age, gender, race, blood pressure (systolic and diastolic), diabetic status, smoking status, various cholesterol lab values, and on certain current medications (hypertension control, statins, or aspirin). This risk prediction tool was built to predict any type of ``hard'' atherosclerotic-based disease, and therefore in addition to predicting future CAD is also predicts future stroke. However, it does not distinguish between risks for these two different outcomes and treats them both as a positive outcome.

\subsubsection{Remote and Dynamic Models}
Remote and telemonitoring studies that use telephones and call-centers as the primary source of data have been used to track HF patients, in the hope of reducing heart failure admissions. These systems are intended to track patient symptoms, including impact of medication, weight gain (as a surrogate for edema), and depression, to identify early signs of decompensation aimed at providing interventions that prevent hospital readmissions in HF patients. In Tele-HF, Krumholz et al. found that a self-report telemonitoring system was not able to reduce readmissions in heart failure patients based upon daily reports of symptoms, medication usage, weight, and depression \cite{RN34}. Ong et al., in the Beat-HF trial, looked to automate some of the data collection surrounding blood pressure and weight, with machine learning risk models to drive interventions, but found similarly that HF readmissions were not reduced \cite{RN33}. Anker et al. surveyed meta-analyses and prospective clinical trials that evaluated the efficacy of telemonitoring in patients with HF \cite{RN32}. They found disagreement between the efficacy of telemonitoring for HF in different types of trials, but stress that the outcomes of telemedicine depend on personalization to the particular patient.

Models that predict risks within varied windows of time have, thus far, been restricted to medical settings. Henry et al. used a rolling model to predict the risk of sepsis in a hospital setting, selecting important features and identifying dynamic risks of sepsis within a single hospital admission \cite{RN127}. Such dynamic models could be adapted to remote and longitudinal settings, but have not done so yet.

\subsubsection{Interpretable Machine Learning}
A recent push in the machine learning field has been to explain predictions provided by deep learning methods that are generally considered black box techniques. Ribiero et al. developed a technique by which local logistic regression models are able to identify the reasons a particular prediction is made based upon the variables that generated the prediction of that element and similar model elements \cite{RN128}. This work demonstrates model interpretability, which comes naturally in CNN deep learning models that can visualize data in intermediate models but becomes much more complicated in time-series based models such as LSTMs. Additional machine learning techniques look to automatically cluster patients and explain the phenotype discovery \cite{RN129}, while also learning to predict multiple outcomes at the same time across different patient types \cite{RN130}, but work on explaining the findings remains in preliminary stages \cite{RN131}.

Interpretability also indicates the confidence in estimations, and understanding what data helps and hurts the predictive accuracy of techniques. In work aimed at improving real-time context and activity detection, Ardywibowo et al. evaluated selected sensors to improve HAR with constraints on the types of sensors and the power those sensors consume \cite{RN132}. Work by Solis et al. use the idea of uncertainty quantification in order to direct users to gather more data in real-time, diet logging settings \cite{RN133}. Uncertainty quantification is an emerging field of interpretable machine learning that has the ability to guide confidence in predictions collected as well as suggest additional data that patients and clinicians should consider collecting.

\subsection{Gaps}
Existing models for predicting risk in cardiovascular conditions rely on sparse data that are measured on rare occasions. Many parameters are trivial to measure (age, gender), and many parameters are Boolean values relating to history. In comparison to the data produced continuous monitoring systems, these data are sparse and likely overly-simplistic. There are two chief ways in which the limitations posed by this sparsity of data can be overcome with richer data: existing models can be updated to include richer data sources, and richer data sources can be analyzed for anomaly detection and rare event detection.

The following gaps remain in developing personalized analytic models based upon the remote sensing data gathered:
\begin{itemize}
    \item Integration of sensing data with acute care data and outcomes for robust risk prediction models.
    \item Development of dynamic models that are flexible to the types of data collected, the windows of data collected, and the changing in patient condition throughout observation.
    \item Interpretable machine learning to explain the predictions of these complex models, and help guide clinical decision making, including identifying similar patients and explaining potentially new phenotypes that might be discovered.
\end{itemize}

\subsection{Opportunities}
Existing models to quantify disease state and future disease risk could be improved through the implementation of richer data sources into the mode. The NYHA Functional Classification of HF relies in part on physical activity levels. The levels are subjective, with definitions in part of ``no limitation of physical activity'' (class I), ``slight limitation of physical activity'' (class II), ``marked limitation of physical activity'' (class III), and ``unable to carry on any physical activity'' (class IV). These classes are inherently subjective, and therefore susceptible to variability between patients with the same underlying disease state. Augmenting this classification with patterns detected from signals such as HAR and effort involved in activity (such as via heartrate monitoring) would allow for objective measurements from beyond the limited scope of direct patient-physician contact. The increase of objective measurements would likely lead to updates to existing models and better information to aid in making clinical decisions.

Existing models could also be improved by the detection of rare or uncommon events. For instance, when a patient presents with AFib, the duration of the AFib is typically unknown. As discussed above, the CHA2DS2-VASc score can aid physicians in predicting stroke and the appropriateness of implementing oral anticoagulation therapy. However, the parameters which contribute to the CHA2DS2-VASc score are simplistically sparse. Age and gender are (for cardiac risk purposes) nonmodifiable risk factors. Each of the other parameters are positive if the patient has ever had the given event once in their life: HF, hypertension, stroke/TIA/thromboembolism, vascular disease, or diabetes. It stands to reason that this model may be improved from richer data, such as the pattern or frequency with which the patient experiences episodes of AFib. Addition of this richer data to the model could potentially result in a model which is better able to discriminate between those at risk of stroke and those at lower risk of stroke, allowing for more appropriate and judicious use of oral anticoagulants.

The emergence of new sensing and internet of things (IoT) technologies creates a need for new models to incorporate new data for better prediction and understanding of disease states. The drastic increase in technology such as smartphones and smartwatches allows for new rich data sources, and also creates a need for the utilization of these data sources. Recently, smartwatches have been adapted to detect conditions such as AFib \cite{RN22}. Further work should look at implementing these new modalities into longitudinal risk models. For instance, HAR recognition could be implemented as a parameter in monitoring activity tolerance in patients with HF. This could supplement the existing subjective measures of heart failure with newer objective measures.

Ultimately, data from new rich data sources is only valuable so far as it contributes to improving the quality of patient healthcare. In order for this contribution to take place, models must generate actionable feedback that can be used for informed clinical decision making. Rather than presenting modeling through a black box approach where data is supplied to the model and an answer is returned, it is desirable that the reasoning behind the risk score is understandable. If a model is interpretable, then the factors leading to a given score can be understood and interventions made to address the risk and to improve patient outcomes. Additionally, the greater the interpretability of a model, the more information that the physician and patient are able to have about the overall disease state. As this information is understood by the physician and the patient, it can be used to better inform and guide care. As a result, the following opportunities exist for immediate and impactful machine learning research:
\begin{itemize}
    \item Machine Learning models with cross-sectional and time-series data: Integration of sensing data with acute care data and outcomes for robust risk prediction models.
    \item Development of dynamic models that are flexible to the types of data collected, the windows of data collected, and the changing in patient condition throughout observation.
    \item Interpretable machine learning to explain the predictions of these complex models, and help guide clinical decision making, including identifying similar patients and explaining potentially new phenotypes that might be discovered.
    \item Transfer Learning: transfer learning techniques will be able to take developed models and adapt to a variety of signals captured, a variety of patients modeled, or a combination therein, improving the flexibility of any analytic techniques developed to advance the prior three opportunities.
\end{itemize}

\section{Discussion and Conclusion}

We surveyed the field of sensing technologies and machine learning analytics that exist in the field of remote monitoring for cardiovascular disorders. Through the evaluation of these sensing modalities and machine learning techniques, we highlighted the potential for addressing three critical areas of need for care in patients diagnosed with heart failure, coronary artery disease, and stroke: 1) the need for sensing technology that can track longitudinal trends in signs and symptoms of the cardiovascular disorder despite potentially infrequent, noisy, or missing data measurements; 2) the need for new analytic techniques that model data captured in a longitudinal, continual fashion to aid in the development of new risk prediction techniques and in tracking disease progression; and 3) the need for machine learning techniques that are personalized and interpretable, allowing for advancements in shared clinical decision making. We highlight these needs based upon the current state-of-the-art in smart health technologies and analytics and discuss the ample opportunities that exist in addressing all three needs in the development of smart health technologies and analytics applied to the field of cardiovascular disorders and care. Whereas the progression of smart health technologies in these needs has demonstrated success in fields such as HAR and physical disorder monitoring, the opportunities for addressing cardiovascular care are many.

These cardiovascular disorders are often very complex conditions characterized by multiple changes in a patient, many of which are slow and difficult to notice. However, systems could be built to take into account and monitor many different changes in order to track disease progression and to allow clinical decisions to be made before rapid decompensations. As disease progresses, regular monitoring of heart sounds could be used in order to track heart remodeling. Instead of noticing these sounds in an acute care visit, computer-aided auscultation though wearable electronic stethoscopes could allow for earlier detection. Quantitative edema tracking would allow for monitoring functional changes within the heart. As pulmonary edema increases, clinicians are able to tell that left heart function is decreasing. Changes in ECG signals may indicate progression of CAD disorders, that may result in additional patient pain, prior to leading to heart attack. As a patient's condition worsens, they may gradually lose the stamina to walk certain distances or to perform a certain amount of activity, demonstrating changes in physical activity capacity, respiratory rate, or sleep quality. These changes may be so gradual that patients may not notice them. Instead, new analytics for progression could instead build activity recognition into the modeling to understand slow changes in baseline function. In this, departures from a patient's baseline level of activity would be significant and could be useful information for guiding clinical care. Similarly, alterations in blood flow may lead to changes in urination habits. As blood flow to the kidneys might be restricted during the day and increased at night due to postural changes, the kidneys will produce more urine at night. Tracking frequency of nocturnal urination could provide more clues as to overall health. Note that this will be sensitive, but not specific for heart failure. Additional measurement of hemodynamic characteristics, such as hypertension, may show treatment effectiveness and better guide the improvement of factors that would lead to conditions such as stroke. In total, a comprehensive system to track the progression of cardiovascular disorders should incorporate a body of integrated sensors, capture this data over longitudinal periods of time, and as a result, enable new advancements in machine learning techniques that can make best use of this data to help guide patient and clinician alike in improving patient care.

This survey highlighted the needs in developing smart health applications to treat HF, CAD, and stroke, and the risk factors associated with them. It reviewed the existing technologies, highlighting the current gaps in solutions presented for those needs. Finally, it presented a series of opportunities, including advanced analytic techniques to be developed once new sensing solutions are available that can guide impactful changes in the way patients with cardiovascular disorders are cared for.

\begin{acks}
This work was supported in part by the National Institutes of Health under grant 1R01EB028106-01. Any opinions, findings, conclusions, or recommendations expressed in this material are those of the authors and do not necessarily reflect the views of the funding organizations.
\end{acks}

\bibliographystyle{ACM-Reference-Format}
\bibliography{cardss_bib_aug_7}


\begin{thebibliography}{115}


\ifx \showCODEN    \undefined \def \showCODEN     #1{\unskip}     \fi
\ifx \showDOI      \undefined \def \showDOI       #1{#1}\fi
\ifx \showISBNx    \undefined \def \showISBNx     #1{\unskip}     \fi
\ifx \showISBNxiii \undefined \def \showISBNxiii  #1{\unskip}     \fi
\ifx \showISSN     \undefined \def \showISSN      #1{\unskip}     \fi
\ifx \showLCCN     \undefined \def \showLCCN      #1{\unskip}     \fi
\ifx \shownote     \undefined \def \shownote      #1{#1}          \fi
\ifx \showarticletitle \undefined \def \showarticletitle #1{#1}   \fi
\ifx \showURL      \undefined \def \showURL       {\relax}        \fi
\providecommand\bibfield[2]{#2}
\providecommand\bibinfo[2]{#2}
\providecommand\natexlab[1]{#1}
\providecommand\showeprint[2][]{arXiv:#2}

\bibitem[\protect\citeauthoryear{Abtahi, Gyllinsky, Paesang, Barlow, Constant,
  Gomes, Tully, D'Andrea, and Mankodiya}{Abtahi et~al\mbox{.}}{2018}]%
        {RN117}
\bibfield{author}{\bibinfo{person}{Mohammadreza Abtahi},
  \bibinfo{person}{Joshua~V Gyllinsky}, \bibinfo{person}{Brandon Paesang},
  \bibinfo{person}{Scott Barlow}, \bibinfo{person}{Matthew Constant},
  \bibinfo{person}{Nicholas Gomes}, \bibinfo{person}{Oliver Tully},
  \bibinfo{person}{Susan~E D'Andrea}, {and} \bibinfo{person}{Kunal Mankodiya}.}
  \bibinfo{year}{2018}\natexlab{}.
\newblock \showarticletitle{MagicSox: An E-textile IoT system to quantify gait
  abnormalities}.
\newblock \bibinfo{journal}{\emph{Smart Health}}  \bibinfo{volume}{5}
  (\bibinfo{year}{2018}), \bibinfo{pages}{4--14}.
\newblock
\showISSN{2352-6483}


\bibitem[\protect\citeauthoryear{Ahmad, Pencina, Schulte, O'Brien, Whellan,
  Piña, Kitzman, Lee, O'Connor, and Felker}{Ahmad et~al\mbox{.}}{2014}]%
        {RN123}
\bibfield{author}{\bibinfo{person}{Tariq Ahmad}, \bibinfo{person}{Michael~J
  Pencina}, \bibinfo{person}{Phillip~J Schulte}, \bibinfo{person}{Emily
  O'Brien}, \bibinfo{person}{David~J Whellan}, \bibinfo{person}{Ileana~L
  Piña}, \bibinfo{person}{Dalane~W Kitzman}, \bibinfo{person}{Kerry~L Lee},
  \bibinfo{person}{Christopher~M O'Connor}, {and} \bibinfo{person}{G~Michael
  Felker}.} \bibinfo{year}{2014}\natexlab{}.
\newblock \showarticletitle{Clinical implications of chronic heart failure
  phenotypes defined by cluster analysis}.
\newblock \bibinfo{journal}{\emph{Journal of the American College of
  Cardiology}} \bibinfo{volume}{64}, \bibinfo{number}{17}
  (\bibinfo{year}{2014}), \bibinfo{pages}{1765--1774}.
\newblock
\showISSN{0735-1097}
\urldef\tempurl%
\url{https://doi.org/10.1016/j.jacc.2014.07.979}
\showDOI{\tempurl}


\bibitem[\protect\citeauthoryear{Amelard, Hughson, Greaves, Pfisterer, Leung,
  Clausi, and Wong}{Amelard et~al\mbox{.}}{2017}]%
        {RN35}
\bibfield{author}{\bibinfo{person}{Robert Amelard}, \bibinfo{person}{Richard~L
  Hughson}, \bibinfo{person}{Danielle~K Greaves}, \bibinfo{person}{Kaylen~J
  Pfisterer}, \bibinfo{person}{Jason Leung}, \bibinfo{person}{David~A Clausi},
  {and} \bibinfo{person}{Alexander Wong}.} \bibinfo{year}{2017}\natexlab{}.
\newblock \showarticletitle{Non-contact hemodynamic imaging reveals the jugular
  venous pulse waveform}.
\newblock \bibinfo{journal}{\emph{Scientific reports}}  \bibinfo{volume}{7}
  (\bibinfo{year}{2017}), \bibinfo{pages}{40150}.
\newblock
\showISSN{2045-2322}
\urldef\tempurl%
\url{https://doi.org/10.1038/srep40150}
\showDOI{\tempurl}


\bibitem[\protect\citeauthoryear{Andersen, Peimankar, and
  Puthusserypady}{Andersen et~al\mbox{.}}{2019}]%
        {RN111}
\bibfield{author}{\bibinfo{person}{Rasmus~S Andersen},
  \bibinfo{person}{Abdolrahman Peimankar}, {and} \bibinfo{person}{Sadasivan
  Puthusserypady}.} \bibinfo{year}{2019}\natexlab{}.
\newblock \showarticletitle{A deep learning approach for real-time detection of
  atrial fibrillation}.
\newblock \bibinfo{journal}{\emph{Expert Systems with Applications}}
  \bibinfo{volume}{115} (\bibinfo{year}{2019}), \bibinfo{pages}{465--473}.
\newblock
\showISSN{0957-4174}


\bibitem[\protect\citeauthoryear{Anker, Koehler, and Abraham}{Anker
  et~al\mbox{.}}{2011}]%
        {RN32}
\bibfield{author}{\bibinfo{person}{Stefan~D Anker}, \bibinfo{person}{Friedrich
  Koehler}, {and} \bibinfo{person}{William~T Abraham}.}
  \bibinfo{year}{2011}\natexlab{}.
\newblock \showarticletitle{Telemedicine and remote management of patients with
  heart failure}.
\newblock \bibinfo{journal}{\emph{The Lancet}} \bibinfo{volume}{378},
  \bibinfo{number}{9792} (\bibinfo{year}{2011}), \bibinfo{pages}{731--739}.
\newblock
\showISSN{0140-6736}
\urldef\tempurl%
\url{https://doi.org/10.1016/S0140-6736(11)61229-4}
\showDOI{\tempurl}


\bibitem[\protect\citeauthoryear{Ardywibowo, Zhao, Wang, Mortazavi, Huang, and
  Qian}{Ardywibowo et~al\mbox{.}}{2019}]%
        {RN132}
\bibfield{author}{\bibinfo{person}{Randy Ardywibowo}, \bibinfo{person}{Guang
  Zhao}, \bibinfo{person}{Zhangyang Wang}, \bibinfo{person}{Bobak Mortazavi},
  \bibinfo{person}{Shuai Huang}, {and} \bibinfo{person}{Xiaoning Qian}.}
  \bibinfo{year}{2019}\natexlab{}.
\newblock \showarticletitle{Adaptive Activity Monitoring with Uncertainty
  Quantification in Switching Gaussian Process Models}. In
  \bibinfo{booktitle}{\emph{The 22nd International Conference on Artificial
  Intelligence and Statistics}}. \bibinfo{pages}{266--275}.
\newblock


\bibitem[\protect\citeauthoryear{Assaad, Kratzert, Shelley, Friedman, and
  Perrino~Jr}{Assaad et~al\mbox{.}}{2018}]%
        {RN46}
\bibfield{author}{\bibinfo{person}{Sherif Assaad}, \bibinfo{person}{Wolf~B
  Kratzert}, \bibinfo{person}{Benjamin Shelley}, \bibinfo{person}{Malcolm~B
  Friedman}, {and} \bibinfo{person}{Albert Perrino~Jr}.}
  \bibinfo{year}{2018}\natexlab{}.
\newblock \showarticletitle{Assessment of pulmonary edema: principles and
  practice}.
\newblock \bibinfo{journal}{\emph{Journal of cardiothoracic and vascular
  anesthesia}} \bibinfo{volume}{32}, \bibinfo{number}{2}
  (\bibinfo{year}{2018}), \bibinfo{pages}{901--914}.
\newblock
\showISSN{1053-0770}
\urldef\tempurl%
\url{https://doi.org/10.1053/j.jvca.2017.08.028}
\showDOI{\tempurl}


\bibitem[\protect\citeauthoryear{Avci, Bosch, Marin-Perianu, Marin-Perianu, and
  Havinga}{Avci et~al\mbox{.}}{2010}]%
        {RN73}
\bibfield{author}{\bibinfo{person}{Akin Avci}, \bibinfo{person}{Stephan Bosch},
  \bibinfo{person}{Mihai Marin-Perianu}, \bibinfo{person}{Raluca
  Marin-Perianu}, {and} \bibinfo{person}{Paul Havinga}.}
  \bibinfo{year}{2010}\natexlab{}.
\newblock \showarticletitle{Activity recognition using inertial sensing for
  healthcare, wellbeing and sports applications: A survey}. In
  \bibinfo{booktitle}{\emph{23th International conference on architecture of
  computing systems 2010}}. \bibinfo{publisher}{VDE}, \bibinfo{pages}{1--10}.
\newblock
\showISBNx{380073222X}


\bibitem[\protect\citeauthoryear{Babbs}{Babbs}{2012}]%
        {RN50}
\bibfield{author}{\bibinfo{person}{Charles~F Babbs}.}
  \bibinfo{year}{2012}\natexlab{}.
\newblock \showarticletitle{Oscillometric measurement of systolic and diastolic
  blood pressures validated in a physiologic mathematical model}.
\newblock \bibinfo{journal}{\emph{Biomedical engineering online}}
  \bibinfo{volume}{11}, \bibinfo{number}{1} (\bibinfo{year}{2012}),
  \bibinfo{pages}{56}.
\newblock
\showISSN{1475-925X}
\urldef\tempurl%
\url{https://doi.org/10.1186/1475-925X-11-56}
\showDOI{\tempurl}


\bibitem[\protect\citeauthoryear{Baker, Einstadter, Husak, and Cebul}{Baker
  et~al\mbox{.}}{2004}]%
        {RN70}
\bibfield{author}{\bibinfo{person}{David~W Baker}, \bibinfo{person}{Doug
  Einstadter}, \bibinfo{person}{Scott~S Husak}, {and}
  \bibinfo{person}{Randall~D Cebul}.} \bibinfo{year}{2004}\natexlab{}.
\newblock \showarticletitle{Trends in postdischarge mortality and readmissions:
  has length of stay declined too far?}
\newblock \bibinfo{journal}{\emph{Archives of internal medicine}}
  \bibinfo{volume}{164}, \bibinfo{number}{5} (\bibinfo{year}{2004}),
  \bibinfo{pages}{538--544}.
\newblock
\showISSN{0003-9926}


\bibitem[\protect\citeauthoryear{Baloglu, Talo, Yildirim, San~Tan, and
  Acharya}{Baloglu et~al\mbox{.}}{2019}]%
        {RN112}
\bibfield{author}{\bibinfo{person}{Ulas~Baran Baloglu},
  \bibinfo{person}{Muhammed Talo}, \bibinfo{person}{Ozal Yildirim},
  \bibinfo{person}{Ru San~Tan}, {and} \bibinfo{person}{U~Rajendra Acharya}.}
  \bibinfo{year}{2019}\natexlab{}.
\newblock \showarticletitle{Classification of myocardial infarction with
  multi-lead ECG signals and deep CNN}.
\newblock \bibinfo{journal}{\emph{Pattern Recognition Letters}}
  \bibinfo{volume}{122} (\bibinfo{year}{2019}), \bibinfo{pages}{23--30}.
\newblock
\showISSN{0167-8655}


\bibitem[\protect\citeauthoryear{Benjamin, Muntner, and Bittencourt}{Benjamin
  et~al\mbox{.}}{2019}]%
        {RN1}
\bibfield{author}{\bibinfo{person}{Emelia~J Benjamin}, \bibinfo{person}{Paul
  Muntner}, {and} \bibinfo{person}{M{\'a}rcio~Sommer Bittencourt}.}
  \bibinfo{year}{2019}\natexlab{}.
\newblock \showarticletitle{Heart disease and stroke statistics-2019 update: A
  report from the American Heart Association}.
\newblock \bibinfo{journal}{\emph{Circulation}} \bibinfo{volume}{139},
  \bibinfo{number}{10} (\bibinfo{year}{2019}), \bibinfo{pages}{e56--e528}.
\newblock
\showISSN{1524-4539 (Electronic) 0009-7322 (Linking)}
\urldef\tempurl%
\url{https://doi.org/10.1161/CIR.0000000000000659}
\showDOI{\tempurl}


\bibitem[\protect\citeauthoryear{Bennett, Savaglio, Lu, Massey, Wang, Wu, and
  Jafari}{Bennett et~al\mbox{.}}{2014}]%
        {RN39}
\bibfield{author}{\bibinfo{person}{Terrell~R Bennett}, \bibinfo{person}{Claudio
  Savaglio}, \bibinfo{person}{David Lu}, \bibinfo{person}{Hunter Massey},
  \bibinfo{person}{Xianan Wang}, \bibinfo{person}{Jian Wu}, {and}
  \bibinfo{person}{Roozbeh Jafari}.} \bibinfo{year}{2014}\natexlab{}.
\newblock \showarticletitle{Motionsynthesis toolset (most): a toolset for human
  motion data synthesis and validation}. In
  \bibinfo{booktitle}{\emph{Proceedings of the 4th ACM MobiHoc workshop on
  Pervasive wireless healthcare}}. \bibinfo{publisher}{ACM},
  \bibinfo{pages}{25--30}.
\newblock
\showISBNx{1450329837}


\bibitem[\protect\citeauthoryear{Bennis, van Pul, van~den Bogaart, Andriessen,
  Kramer, and Delhaas}{Bennis et~al\mbox{.}}{2019}]%
        {RN90}
\bibfield{author}{\bibinfo{person}{Frank~C Bennis}, \bibinfo{person}{Carola van
  Pul}, \bibinfo{person}{Jarno~JL van~den Bogaart}, \bibinfo{person}{Peter
  Andriessen}, \bibinfo{person}{Boris~W Kramer}, {and} \bibinfo{person}{Tammo
  Delhaas}.} \bibinfo{year}{2019}\natexlab{}.
\newblock \showarticletitle{Artifacts in pulse transit time measurements using
  standard patient monitoring equipment}.
\newblock \bibinfo{journal}{\emph{PloS one}} \bibinfo{volume}{14},
  \bibinfo{number}{6} (\bibinfo{year}{2019}), \bibinfo{pages}{e0218784}.
\newblock
\showISSN{1932-6203}


\bibitem[\protect\citeauthoryear{Bhuyan, Choe, Lee, Cristman, Oralkan, and
  Khuri-Yakub}{Bhuyan et~al\mbox{.}}{2011}]%
        {RN45}
\bibfield{author}{\bibinfo{person}{Anshuman Bhuyan}, \bibinfo{person}{Jung~Woo
  Choe}, \bibinfo{person}{Byung~Chul Lee}, \bibinfo{person}{Paul Cristman},
  \bibinfo{person}{{\"O}mer Oralkan}, {and} \bibinfo{person}{Butrus~T
  Khuri-Yakub}.} \bibinfo{year}{2011}\natexlab{}.
\newblock \showarticletitle{Miniaturized, wearable, ultrasound probe for
  on-demand ultrasound screening}. In \bibinfo{booktitle}{\emph{2011 IEEE
  International Ultrasonics Symposium}}. \bibinfo{publisher}{IEEE},
  \bibinfo{pages}{1060--1063}.
\newblock
\showISBNx{1457712520}


\bibitem[\protect\citeauthoryear{Bumgarner, Lambert, Hussein, Cantillon,
  Baranowski, Wolski, Lindsay, Wazni, and Tarakji}{Bumgarner
  et~al\mbox{.}}{2018}]%
        {RN22}
\bibfield{author}{\bibinfo{person}{Joseph~M Bumgarner},
  \bibinfo{person}{Cameron~T Lambert}, \bibinfo{person}{Ayman~A Hussein},
  \bibinfo{person}{Daniel~J Cantillon}, \bibinfo{person}{Bryan Baranowski},
  \bibinfo{person}{Kathy Wolski}, \bibinfo{person}{Bruce~D Lindsay},
  \bibinfo{person}{Oussama~M Wazni}, {and} \bibinfo{person}{Khaldoun~G
  Tarakji}.} \bibinfo{year}{2018}\natexlab{}.
\newblock \showarticletitle{Smartwatch algorithm for automated detection of
  atrial fibrillation}.
\newblock \bibinfo{journal}{\emph{Journal of the American College of
  Cardiology}} \bibinfo{volume}{71}, \bibinfo{number}{21}
  (\bibinfo{year}{2018}), \bibinfo{pages}{2381--2388}.
\newblock
\showISSN{0735-1097}
\urldef\tempurl%
\url{https://doi.org/10.1016/j.jacc.2018.03.003}
\showDOI{\tempurl}


\bibitem[\protect\citeauthoryear{Chaudhry, Mattera, Curtis, Spertus, Herrin,
  Lin, Phillips, Hodshon, Cooper, and Krumholz}{Chaudhry et~al\mbox{.}}{2010}]%
        {RN71}
\bibfield{author}{\bibinfo{person}{Sarwat~I Chaudhry},
  \bibinfo{person}{Jennifer~A Mattera}, \bibinfo{person}{Jeptha~P Curtis},
  \bibinfo{person}{John~A Spertus}, \bibinfo{person}{Jeph Herrin},
  \bibinfo{person}{Zhenqiu Lin}, \bibinfo{person}{Christopher~O Phillips},
  \bibinfo{person}{Beth~V Hodshon}, \bibinfo{person}{Lawton~S Cooper}, {and}
  \bibinfo{person}{Harlan~M Krumholz}.} \bibinfo{year}{2010}\natexlab{}.
\newblock \showarticletitle{Telemonitoring in patients with heart failure}.
\newblock \bibinfo{journal}{\emph{New England Journal of Medicine}}
  \bibinfo{volume}{363}, \bibinfo{number}{24} (\bibinfo{year}{2010}),
  \bibinfo{pages}{2301--2309}.
\newblock
\showISSN{0028-4793}
\urldef\tempurl%
\url{https://doi.org/10.1056/NEJMoa1010029}
\showDOI{\tempurl}


\bibitem[\protect\citeauthoryear{Chen, Jafari, and Kehtarnavaz}{Chen
  et~al\mbox{.}}{2017a}]%
        {RN74}
\bibfield{author}{\bibinfo{person}{Chen Chen}, \bibinfo{person}{Roozbeh
  Jafari}, {and} \bibinfo{person}{Nasser Kehtarnavaz}.}
  \bibinfo{year}{2017}\natexlab{a}.
\newblock \showarticletitle{A survey of depth and inertial sensor fusion for
  human action recognition}.
\newblock \bibinfo{journal}{\emph{Multimedia Tools and Applications}}
  \bibinfo{volume}{76}, \bibinfo{number}{3} (\bibinfo{year}{2017}),
  \bibinfo{pages}{4405--4425}.
\newblock
\showISSN{1380-7501}


\bibitem[\protect\citeauthoryear{Chen, Peng, and Razi}{Chen
  et~al\mbox{.}}{2017b}]%
        {RN52}
\bibfield{author}{\bibinfo{person}{Jiaming Chen}, \bibinfo{person}{Han Peng},
  {and} \bibinfo{person}{Abolfazl Razi}.} \bibinfo{year}{2017}\natexlab{b}.
\newblock \showarticletitle{Remote ECG monitoring kit to predict
  patient-specific heart abnormalities}.
\newblock \bibinfo{journal}{\emph{Journal of Systemics, Cybernetics and
  Informatics}} \bibinfo{volume}{15}, \bibinfo{number}{4}
  (\bibinfo{year}{2017}), \bibinfo{pages}{82--89}.
\newblock


\bibitem[\protect\citeauthoryear{Conn and O'Keefe}{Conn and O'Keefe}{2009}]%
        {RN10}
\bibfield{author}{\bibinfo{person}{Robert~D Conn} {and}
  \bibinfo{person}{James~H O'Keefe}.} \bibinfo{year}{2009}\natexlab{}.
\newblock \showarticletitle{Cardiac physical diagnosis in the digital age: an
  important but increasingly neglected skill (from stethoscopes to
  microchips)}.
\newblock \bibinfo{journal}{\emph{The American journal of cardiology}}
  \bibinfo{volume}{104}, \bibinfo{number}{4} (\bibinfo{year}{2009}),
  \bibinfo{pages}{590--595}.
\newblock
\showISSN{0002-9149}
\urldef\tempurl%
\url{https://doi.org/10.1016/j.amjcard.2009.04.030}
\showDOI{\tempurl}


\bibitem[\protect\citeauthoryear{Dai, Zhang, Liu, Ding, and Zhao}{Dai
  et~al\mbox{.}}{2016}]%
        {RN91}
\bibfield{author}{\bibinfo{person}{Wen-Xuan Dai}, \bibinfo{person}{Yuan-Ting
  Zhang}, \bibinfo{person}{Jing Liu}, \bibinfo{person}{Xiao-Rong Ding}, {and}
  \bibinfo{person}{Ni Zhao}.} \bibinfo{year}{2016}\natexlab{}.
\newblock \showarticletitle{Dual-modality arterial pulse monitoring system for
  continuous blood pressure measurement}. In \bibinfo{booktitle}{\emph{2016
  38th Annual International Conference of the IEEE Engineering in Medicine and
  Biology Society (EMBC)}}. \bibinfo{publisher}{IEEE},
  \bibinfo{pages}{5773--5776}.
\newblock
\showISBNx{1457702207}


\bibitem[\protect\citeauthoryear{de~Denus, O'Meara, Desai, Claggett, Lewis,
  Leclair, Jutras, Lavoie, Solomon, and Pitt}{de~Denus et~al\mbox{.}}{2017}]%
        {RN125}
\bibfield{author}{\bibinfo{person}{Simon de Denus}, \bibinfo{person}{Eileen
  O'Meara}, \bibinfo{person}{Akshay~S Desai}, \bibinfo{person}{Brian Claggett},
  \bibinfo{person}{Eldrin~F Lewis}, \bibinfo{person}{Gr{\'e}goire Leclair},
  \bibinfo{person}{Martin Jutras}, \bibinfo{person}{Joël Lavoie},
  \bibinfo{person}{Scott~D Solomon}, {and} \bibinfo{person}{Bertram Pitt}.}
  \bibinfo{year}{2017}\natexlab{}.
\newblock \showarticletitle{Spironolactone metabolites in TOPCAT—new insights
  into regional variation}.
\newblock \bibinfo{journal}{\emph{New England Journal of Medicine}}
  \bibinfo{volume}{376}, \bibinfo{number}{17} (\bibinfo{year}{2017}),
  \bibinfo{pages}{1690--1692}.
\newblock
\showISSN{0028-4793}


\bibitem[\protect\citeauthoryear{Dobkin and Martinez}{Dobkin and
  Martinez}{2018}]%
        {RN79}
\bibfield{author}{\bibinfo{person}{Bruce~H Dobkin} {and}
  \bibinfo{person}{Clarisa Martinez}.} \bibinfo{year}{2018}\natexlab{}.
\newblock \showarticletitle{Wearable Sensors to Monitor, Enable Feedback, and
  Measure Outcomes of Activity and Practice}.
\newblock \bibinfo{journal}{\emph{Current neurology and neuroscience reports}}
  \bibinfo{volume}{18}, \bibinfo{number}{12} (\bibinfo{year}{2018}),
  \bibinfo{pages}{87}.
\newblock
\showISSN{1528-4042}
\urldef\tempurl%
\url{https://doi.org/10.1007/s11910-018-0896-5}
\showDOI{\tempurl}


\bibitem[\protect\citeauthoryear{Elgendi, Bobhate, Jain, Rutledge, Coe, Zemp,
  Schuurmans, and Adatia}{Elgendi et~al\mbox{.}}{2014}]%
        {RN63}
\bibfield{author}{\bibinfo{person}{Mohamed Elgendi}, \bibinfo{person}{Prashant
  Bobhate}, \bibinfo{person}{Shreepal Jain}, \bibinfo{person}{Jennifer
  Rutledge}, \bibinfo{person}{James~Y Coe}, \bibinfo{person}{Roger Zemp},
  \bibinfo{person}{Dale Schuurmans}, {and} \bibinfo{person}{Ian Adatia}.}
  \bibinfo{year}{2014}\natexlab{}.
\newblock \showarticletitle{Time-domain analysis of heart sound intensity in
  children with and without pulmonary artery hypertension: a pilot study using
  a digital stethoscope}.
\newblock \bibinfo{journal}{\emph{Pulmonary circulation}} \bibinfo{volume}{4},
  \bibinfo{number}{4} (\bibinfo{year}{2014}), \bibinfo{pages}{685--695}.
\newblock
\showISSN{2045-8932}


\bibitem[\protect\citeauthoryear{Eskofier, Lee, Baron, Simon, Martindale,
  Ga{\ss}ner, and Klucken}{Eskofier et~al\mbox{.}}{2017}]%
        {RN75}
\bibfield{author}{\bibinfo{person}{Bjoern Eskofier}, \bibinfo{person}{Sunghoon
  Lee}, \bibinfo{person}{Manuela Baron}, \bibinfo{person}{Andr{\'e} Simon},
  \bibinfo{person}{Christine Martindale}, \bibinfo{person}{Heiko Ga{\ss}ner},
  {and} \bibinfo{person}{Jochen Klucken}.} \bibinfo{year}{2017}\natexlab{}.
\newblock \showarticletitle{An overview of smart shoes in the internet of
  health things: gait and mobility assessment in health promotion and disease
  monitoring}.
\newblock \bibinfo{journal}{\emph{Applied Sciences}} \bibinfo{volume}{7},
  \bibinfo{number}{10} (\bibinfo{year}{2017}), \bibinfo{pages}{986}.
\newblock


\bibitem[\protect\citeauthoryear{Fallahzadeh, Pedram, and
  Ghasemzadeh}{Fallahzadeh et~al\mbox{.}}{2016}]%
        {RN41}
\bibfield{author}{\bibinfo{person}{Ramin Fallahzadeh}, \bibinfo{person}{Mahdi
  Pedram}, {and} \bibinfo{person}{Hassan Ghasemzadeh}.}
  \bibinfo{year}{2016}\natexlab{}.
\newblock \showarticletitle{Smartsock: A wearable platform for context-aware
  assessment of ankle edema}. In \bibinfo{booktitle}{\emph{2016 38th Annual
  International Conference of the IEEE Engineering in Medicine and Biology
  Society (EMBC)}}. \bibinfo{publisher}{IEEE}, \bibinfo{pages}{6302--6306}.
\newblock
\showISBNx{1457702207}


\bibitem[\protect\citeauthoryear{Fallahzadeh, Pedram, Saeedi, Sadeghi, Ong, and
  Ghasemzadeh}{Fallahzadeh et~al\mbox{.}}{2015}]%
        {RN98}
\bibfield{author}{\bibinfo{person}{Ramin Fallahzadeh}, \bibinfo{person}{Mahdi
  Pedram}, \bibinfo{person}{Ramyar Saeedi}, \bibinfo{person}{Bahman Sadeghi},
  \bibinfo{person}{Michael Ong}, {and} \bibinfo{person}{Hassan Ghasemzadeh}.}
  \bibinfo{year}{2015}\natexlab{}.
\newblock \showarticletitle{Smart-cuff: A wearable bio-sensing platform with
  activity-sensitive information quality assessment for monitoring ankle
  edema}. In \bibinfo{booktitle}{\emph{2015 IEEE International Conference on
  Pervasive Computing and Communication Workshops (PerCom Workshops)}}.
  \bibinfo{publisher}{IEEE}, \bibinfo{pages}{57--62}.
\newblock
\showISBNx{1479984256}


\bibitem[\protect\citeauthoryear{Fox, Ang, Jaiswal, Pop-Busui, and Wiens}{Fox
  et~al\mbox{.}}{2017}]%
        {RN80}
\bibfield{author}{\bibinfo{person}{Ian Fox}, \bibinfo{person}{Lynn Ang},
  \bibinfo{person}{Mamta Jaiswal}, \bibinfo{person}{Rodica Pop-Busui}, {and}
  \bibinfo{person}{Jenna Wiens}.} \bibinfo{year}{2017}\natexlab{}.
\newblock \showarticletitle{Contextual motifs: Increasing the utility of motifs
  using contextual data}. In \bibinfo{booktitle}{\emph{Proceedings of the 23rd
  ACM SIGKDD International Conference on Knowledge Discovery and Data Mining}}.
  \bibinfo{publisher}{ACM}, \bibinfo{pages}{155--164}.
\newblock
\showISBNx{1450348874}


\bibitem[\protect\citeauthoryear{Fox, Ang, Jaiswal, Pop-Busui, and Wiens}{Fox
  et~al\mbox{.}}{2018}]%
        {RN82}
\bibfield{author}{\bibinfo{person}{Ian Fox}, \bibinfo{person}{Lynn Ang},
  \bibinfo{person}{Mamta Jaiswal}, \bibinfo{person}{Rodica Pop-Busui}, {and}
  \bibinfo{person}{Jenna Wiens}.} \bibinfo{year}{2018}\natexlab{}.
\newblock \showarticletitle{Deep multi-output forecasting: Learning to
  accurately predict blood glucose trajectories}. In
  \bibinfo{booktitle}{\emph{Proceedings of the 24th ACM SIGKDD International
  Conference on Knowledge Discovery \& Data Mining}}. \bibinfo{publisher}{ACM},
  \bibinfo{pages}{1387--1395}.
\newblock
\showISBNx{1450355528}


\bibitem[\protect\citeauthoryear{Gee, Garcia-Olano, Ghosh, and Paydarfar}{Gee
  et~al\mbox{.}}{2019}]%
        {RN131}
\bibfield{author}{\bibinfo{person}{Alan~H Gee}, \bibinfo{person}{Diego
  Garcia-Olano}, \bibinfo{person}{Joydeep Ghosh}, {and} \bibinfo{person}{David
  Paydarfar}.} \bibinfo{year}{2019}\natexlab{}.
\newblock \showarticletitle{Explaining Deep Classification of Time-Series Data
  with Learned Prototypes}.
\newblock \bibinfo{journal}{\emph{arXiv preprint arXiv:1904.08935}}
  (\bibinfo{year}{2019}).
\newblock


\bibitem[\protect\citeauthoryear{Goff, Lloyd-Jones, Bennett, Coady, D'Agostino,
  Gibbons, Greenland, Lackland, Levy, and O'Donnell}{Goff
  et~al\mbox{.}}{2014}]%
        {RN19}
\bibfield{author}{\bibinfo{person}{David~C Goff}, \bibinfo{person}{Donald~M
  Lloyd-Jones}, \bibinfo{person}{Glen Bennett}, \bibinfo{person}{Sean Coady},
  \bibinfo{person}{Ralph~B D'Agostino}, \bibinfo{person}{Raymond Gibbons},
  \bibinfo{person}{Philip Greenland}, \bibinfo{person}{Daniel~T Lackland},
  \bibinfo{person}{Daniel Levy}, {and} \bibinfo{person}{Christopher~J
  O'Donnell}.} \bibinfo{year}{2014}\natexlab{}.
\newblock \showarticletitle{2013 ACC/AHA guideline on the assessment of
  cardiovascular risk: a report of the American College of Cardiology/American
  Heart Association Task Force on Practice Guidelines}.
\newblock \bibinfo{journal}{\emph{Journal of the American College of
  Cardiology}} \bibinfo{volume}{63}, \bibinfo{number}{25 Part B}
  (\bibinfo{year}{2014}), \bibinfo{pages}{2935--2959}.
\newblock
\showISSN{0735-1097}


\bibitem[\protect\citeauthoryear{Group}{Group}{2015}]%
        {RN68}
\bibfield{author}{\bibinfo{person}{SPRINT~Research Group}.}
  \bibinfo{year}{2015}\natexlab{}.
\newblock \showarticletitle{A randomized trial of intensive versus standard
  blood-pressure control}.
\newblock \bibinfo{journal}{\emph{New England Journal of Medicine}}
  \bibinfo{volume}{373}, \bibinfo{number}{22} (\bibinfo{year}{2015}),
  \bibinfo{pages}{2103--2116}.
\newblock


\bibitem[\protect\citeauthoryear{Henderson, He, Malin, Denny, Kho, Ghosh, and
  Ho}{Henderson et~al\mbox{.}}{2018}]%
        {RN129}
\bibfield{author}{\bibinfo{person}{Jette Henderson}, \bibinfo{person}{Huan He},
  \bibinfo{person}{Bradley~A Malin}, \bibinfo{person}{Joshua~C Denny},
  \bibinfo{person}{Abel~N Kho}, \bibinfo{person}{Joydeep Ghosh}, {and}
  \bibinfo{person}{Joyce~C Ho}.} \bibinfo{year}{2018}\natexlab{}.
\newblock \showarticletitle{Phenotyping through Semi-Supervised Tensor
  Factorization (PSST)}. In \bibinfo{booktitle}{\emph{AMIA Annual Symposium
  Proceedings}}, Vol.~\bibinfo{volume}{2018}. \bibinfo{publisher}{American
  Medical Informatics Association}, \bibinfo{pages}{564}.
\newblock


\bibitem[\protect\citeauthoryear{Henry, Hager, Pronovost, and Saria}{Henry
  et~al\mbox{.}}{2015}]%
        {RN127}
\bibfield{author}{\bibinfo{person}{Katharine~E Henry}, \bibinfo{person}{David~N
  Hager}, \bibinfo{person}{Peter~J Pronovost}, {and} \bibinfo{person}{Suchi
  Saria}.} \bibinfo{year}{2015}\natexlab{}.
\newblock \showarticletitle{A targeted real-time early warning score
  (TREWScore) for septic shock}.
\newblock \bibinfo{journal}{\emph{Science translational medicine}}
  \bibinfo{volume}{7}, \bibinfo{number}{299} (\bibinfo{year}{2015}),
  \bibinfo{pages}{299ra122--299ra122}.
\newblock
\showISSN{1946-6234}
\urldef\tempurl%
\url{https://doi.org/10.1126/scitranslmed.aab3719}
\showDOI{\tempurl}


\bibitem[\protect\citeauthoryear{Heran, Chen, Ebrahim, Moxham, Oldridge, Rees,
  Thompson, and Taylor}{Heran et~al\mbox{.}}{2011}]%
        {RN118}
\bibfield{author}{\bibinfo{person}{Balraj~S Heran}, \bibinfo{person}{Jenny~MH
  Chen}, \bibinfo{person}{Shah Ebrahim}, \bibinfo{person}{Tiffany Moxham},
  \bibinfo{person}{Neil Oldridge}, \bibinfo{person}{Karen Rees},
  \bibinfo{person}{David~R Thompson}, {and} \bibinfo{person}{Rod~S Taylor}.}
  \bibinfo{year}{2011}\natexlab{}.
\newblock \showarticletitle{Exercise-based cardiac rehabilitation for coronary
  heart disease}.
\newblock \bibinfo{journal}{\emph{Cochrane database of systematic reviews}}
  \bibinfo{number}{7} (\bibinfo{year}{2011}), \bibinfo{pages}{CD001800}.
\newblock
\showISSN{1465-1858}
\urldef\tempurl%
\url{https://doi.org/10.1002/14651858.CD001800.pub2}
\showDOI{\tempurl}


\bibitem[\protect\citeauthoryear{Hijazi, Page, Kantarci, and Soyata}{Hijazi
  et~al\mbox{.}}{2016}]%
        {RN83}
\bibfield{author}{\bibinfo{person}{Shurouq Hijazi}, \bibinfo{person}{Alex
  Page}, \bibinfo{person}{Burak Kantarci}, {and} \bibinfo{person}{Tolga
  Soyata}.} \bibinfo{year}{2016}\natexlab{}.
\newblock \showarticletitle{Machine learning in cardiac health monitoring and
  decision support}.
\newblock \bibinfo{journal}{\emph{Computer}} \bibinfo{volume}{49},
  \bibinfo{number}{11} (\bibinfo{year}{2016}), \bibinfo{pages}{38--48}.
\newblock
\showISSN{0018-9162}


\bibitem[\protect\citeauthoryear{Hippisley-Cox, Coupland, and
  Brindle}{Hippisley-Cox et~al\mbox{.}}{2013}]%
        {RN18}
\bibfield{author}{\bibinfo{person}{Julia Hippisley-Cox}, \bibinfo{person}{Carol
  Coupland}, {and} \bibinfo{person}{Peter Brindle}.}
  \bibinfo{year}{2013}\natexlab{}.
\newblock \showarticletitle{Derivation and validation of QStroke score for
  predicting risk of ischaemic stroke in primary care and comparison with other
  risk scores: a prospective open cohort study}.
\newblock \bibinfo{journal}{\emph{Bmj}}  \bibinfo{volume}{346}
  (\bibinfo{year}{2013}), \bibinfo{pages}{f2573}.
\newblock
\showISSN{1756-1833}
\urldef\tempurl%
\url{https://doi.org/10.1136/bmj.f2573}
\showDOI{\tempurl}


\bibitem[\protect\citeauthoryear{Hodgkinson, Mant, Martin, Guo, Hobbs, Deeks,
  Heneghan, Roberts, and McManus}{Hodgkinson et~al\mbox{.}}{2011}]%
        {RN49}
\bibfield{author}{\bibinfo{person}{J Hodgkinson}, \bibinfo{person}{J Mant},
  \bibinfo{person}{U Martin}, \bibinfo{person}{B Guo}, \bibinfo{person}{FDR
  Hobbs}, \bibinfo{person}{JJ Deeks}, \bibinfo{person}{C Heneghan},
  \bibinfo{person}{N Roberts}, {and} \bibinfo{person}{RJ McManus}.}
  \bibinfo{year}{2011}\natexlab{}.
\newblock \showarticletitle{Relative effectiveness of clinic and home blood
  pressure monitoring compared with ambulatory blood pressure monitoring in
  diagnosis of hypertension: systematic review}.
\newblock \bibinfo{journal}{\emph{Bmj}}  \bibinfo{volume}{342}
  (\bibinfo{year}{2011}), \bibinfo{pages}{d3621}.
\newblock
\showISSN{0959-8138}
\urldef\tempurl%
\url{https://doi.org/10.1136/bmj.d3621}
\showDOI{\tempurl}


\bibitem[\protect\citeauthoryear{Hoque and Stankovic}{Hoque and
  Stankovic}{2012}]%
        {RN76}
\bibfield{author}{\bibinfo{person}{Enamul Hoque} {and} \bibinfo{person}{John
  Stankovic}.} \bibinfo{year}{2012}\natexlab{}.
\newblock \showarticletitle{AALO: Activity recognition in smart homes using
  Active Learning in the presence of Overlapped activities}. In
  \bibinfo{booktitle}{\emph{2012 6th International Conference on Pervasive
  Computing Technologies for Healthcare (PervasiveHealth) and Workshops}}.
  \bibinfo{publisher}{IEEE}, \bibinfo{pages}{139--146}.
\newblock
\showISBNx{1936968436}


\bibitem[\protect\citeauthoryear{Huynh, Jafari, and Chung}{Huynh
  et~al\mbox{.}}{2018}]%
        {RN93}
\bibfield{author}{\bibinfo{person}{Toan Huynh}, \bibinfo{person}{Roozbeh
  Jafari}, {and} \bibinfo{person}{Wan-Young Chung}.}
  \bibinfo{year}{2018}\natexlab{}.
\newblock \showarticletitle{An Accurate Bioimpedance measurement system for
  blood pressure monitoring}.
\newblock \bibinfo{journal}{\emph{Sensors}} \bibinfo{volume}{18},
  \bibinfo{number}{7} (\bibinfo{year}{2018}), \bibinfo{pages}{2095}.
\newblock


\bibitem[\protect\citeauthoryear{Ibrahim, Akbari, and Jafari}{Ibrahim
  et~al\mbox{.}}{2017}]%
        {RN94}
\bibfield{author}{\bibinfo{person}{Bassem Ibrahim}, \bibinfo{person}{Ali
  Akbari}, {and} \bibinfo{person}{Roozbeh Jafari}.}
  \bibinfo{year}{2017}\natexlab{}.
\newblock \showarticletitle{A novel method for pulse transit time estimation
  using wrist bio-impedance sensing based on a regression model}. In
  \bibinfo{booktitle}{\emph{2017 IEEE Biomedical Circuits and Systems
  Conference (BioCAS)}}. \bibinfo{publisher}{IEEE}, \bibinfo{pages}{1--4}.
\newblock
\showISBNx{1509058036}


\bibitem[\protect\citeauthoryear{Inan, Javaid, Dowling, Ashouri, Etemadi,
  Heller, Roy, and Klein}{Inan et~al\mbox{.}}{2016}]%
        {RN95}
\bibfield{author}{\bibinfo{person}{Omer~T Inan}, \bibinfo{person}{Abdul~Q
  Javaid}, \bibinfo{person}{Sean Dowling}, \bibinfo{person}{Hazar Ashouri},
  \bibinfo{person}{Mozziyar Etemadi}, \bibinfo{person}{James~A Heller},
  \bibinfo{person}{Shuvo Roy}, {and} \bibinfo{person}{Liviu Klein}.}
  \bibinfo{year}{2016}\natexlab{}.
\newblock \showarticletitle{Using ballistocardiography to monitor left
  ventricular function in heart failure patients}.
\newblock \bibinfo{journal}{\emph{Journal of Cardiac Failure}}
  \bibinfo{volume}{22}, \bibinfo{number}{8} (\bibinfo{year}{2016}),
  \bibinfo{pages}{S45}.
\newblock
\showISSN{1071-9164}


\bibitem[\protect\citeauthoryear{Jambukia, Dabhi, and Prajapati}{Jambukia
  et~al\mbox{.}}{2015}]%
        {RN58}
\bibfield{author}{\bibinfo{person}{Shweta~H Jambukia}, \bibinfo{person}{Vipul~K
  Dabhi}, {and} \bibinfo{person}{Harshadkumar~B Prajapati}.}
  \bibinfo{year}{2015}\natexlab{}.
\newblock \showarticletitle{Classification of ECG signals using machine
  learning techniques: A survey}. In \bibinfo{booktitle}{\emph{2015
  International Conference on Advances in Computer Engineering and
  Applications}}. \bibinfo{publisher}{IEEE}, \bibinfo{pages}{714--721}.
\newblock
\showISBNx{146736911X}


\bibitem[\protect\citeauthoryear{Karmali, Davies, Taylor, Beswick, Martin, and
  Ebrahim}{Karmali et~al\mbox{.}}{2014}]%
        {RN120}
\bibfield{author}{\bibinfo{person}{Kunal~N Karmali}, \bibinfo{person}{Philippa
  Davies}, \bibinfo{person}{Fiona Taylor}, \bibinfo{person}{Andrew Beswick},
  \bibinfo{person}{Nicole Martin}, {and} \bibinfo{person}{Shah Ebrahim}.}
  \bibinfo{year}{2014}\natexlab{}.
\newblock \showarticletitle{Promoting patient uptake and adherence in cardiac
  rehabilitation}.
\newblock \bibinfo{journal}{\emph{Cochrane Database of Systematic Reviews}}
  \bibinfo{number}{6} (\bibinfo{year}{2014}), \bibinfo{pages}{CD007131}.
\newblock
\showISSN{1465-1858}
\urldef\tempurl%
\url{https://doi.org/10.1002/14651858.CD007131.pub3}
\showDOI{\tempurl}


\bibitem[\protect\citeauthoryear{Kim, Carek, Inan, Mukkamala, and Hahn}{Kim
  et~al\mbox{.}}{2018}]%
        {RN97}
\bibfield{author}{\bibinfo{person}{Chang-Sei Kim}, \bibinfo{person}{Andrew~M
  Carek}, \bibinfo{person}{Omer~T Inan}, \bibinfo{person}{Ramakrishna
  Mukkamala}, {and} \bibinfo{person}{Jin-Oh Hahn}.}
  \bibinfo{year}{2018}\natexlab{}.
\newblock \showarticletitle{Ballistocardiogram-based approach to cuffless blood
  pressure monitoring: proof of concept and potential challenges}.
\newblock \bibinfo{journal}{\emph{IEEE Transactions on Biomedical Engineering}}
  \bibinfo{volume}{65}, \bibinfo{number}{11} (\bibinfo{year}{2018}),
  \bibinfo{pages}{2384--2391}.
\newblock
\showISSN{0018-9294}
\urldef\tempurl%
\url{https://doi.org/10.1109/TBME.2018.2797239}
\showDOI{\tempurl}


\bibitem[\protect\citeauthoryear{Kim, Ober, McMurtry, Finegan, Inan, Mukkamala,
  and Hahn}{Kim et~al\mbox{.}}{2016}]%
        {RN96}
\bibfield{author}{\bibinfo{person}{Chang-Sei Kim}, \bibinfo{person}{Stephanie~L
  Ober}, \bibinfo{person}{M~Sean McMurtry}, \bibinfo{person}{Barry~A Finegan},
  \bibinfo{person}{Omer~T Inan}, \bibinfo{person}{Ramakrishna Mukkamala}, {and}
  \bibinfo{person}{Jin-Oh Hahn}.} \bibinfo{year}{2016}\natexlab{}.
\newblock \showarticletitle{Ballistocardiogram: Mechanism and potential for
  unobtrusive cardiovascular health monitoring}.
\newblock \bibinfo{journal}{\emph{Scientific reports}}  \bibinfo{volume}{6}
  (\bibinfo{year}{2016}), \bibinfo{pages}{31297}.
\newblock
\showISSN{2045-2322}
\urldef\tempurl%
\url{https://doi.org/10.1038/srep31297}
\showDOI{\tempurl}


\bibitem[\protect\citeauthoryear{Knapp, Cetrullo, Sillars, Lenzo, Davis, and
  Davis}{Knapp et~al\mbox{.}}{2014}]%
        {RN62}
\bibfield{author}{\bibinfo{person}{Arthur Knapp}, \bibinfo{person}{Violetta
  Cetrullo}, \bibinfo{person}{Brett~A Sillars}, \bibinfo{person}{Nat Lenzo},
  \bibinfo{person}{Wendy~A Davis}, {and} \bibinfo{person}{Timothy~ME Davis}.}
  \bibinfo{year}{2014}\natexlab{}.
\newblock \showarticletitle{Carotid artery ultrasonographic assessment in
  patients from the Fremantle Diabetes Study Phase II with carotid bruits
  detected by electronic auscultation}.
\newblock \bibinfo{journal}{\emph{Diabetes technology \& therapeutics}}
  \bibinfo{volume}{16}, \bibinfo{number}{9} (\bibinfo{year}{2014}),
  \bibinfo{pages}{604--610}.
\newblock
\showISSN{1520-9156}
\urldef\tempurl%
\url{https://doi.org/10.1089/dia.2014.0048}
\showDOI{\tempurl}


\bibitem[\protect\citeauthoryear{Krumholz, Chaudhry, Spertus, Mattera, Hodshon,
  and Herrin}{Krumholz et~al\mbox{.}}{2016}]%
        {RN34}
\bibfield{author}{\bibinfo{person}{Harlan~M Krumholz},
  \bibinfo{person}{Sarwat~I Chaudhry}, \bibinfo{person}{John~A Spertus},
  \bibinfo{person}{Jennifer~A Mattera}, \bibinfo{person}{Beth Hodshon}, {and}
  \bibinfo{person}{Jeph Herrin}.} \bibinfo{year}{2016}\natexlab{}.
\newblock \showarticletitle{Do non-clinical factors improve prediction of
  readmission risk?: results from the Tele-HF study}.
\newblock \bibinfo{journal}{\emph{JACC: Heart Failure}} \bibinfo{volume}{4},
  \bibinfo{number}{1} (\bibinfo{year}{2016}), \bibinfo{pages}{12--20}.
\newblock
\showISSN{2213-1779}


\bibitem[\protect\citeauthoryear{Krumholz, Herrin, Miller, Drye, Ling, Han,
  Rapp, Bradley, Nallamothu, and Nsa}{Krumholz et~al\mbox{.}}{2011}]%
        {RN3}
\bibfield{author}{\bibinfo{person}{Harlan~M Krumholz}, \bibinfo{person}{Jeph
  Herrin}, \bibinfo{person}{Lauren~E Miller}, \bibinfo{person}{Elizabeth~E
  Drye}, \bibinfo{person}{Shari~M Ling}, \bibinfo{person}{Lein~F Han},
  \bibinfo{person}{Michael~T Rapp}, \bibinfo{person}{Elizabeth~H Bradley},
  \bibinfo{person}{Brahmajee~K Nallamothu}, {and} \bibinfo{person}{Wato Nsa}.}
  \bibinfo{year}{2011}\natexlab{}.
\newblock \showarticletitle{Improvements in door-to-balloon time in the United
  States, 2005 to 2010}.
\newblock \bibinfo{journal}{\emph{Circulation}} \bibinfo{volume}{124},
  \bibinfo{number}{9} (\bibinfo{year}{2011}), \bibinfo{pages}{1038--1045}.
\newblock
\showISSN{0009-7322}
\urldef\tempurl%
\url{https://doi.org/10.1161/CIRCULATIONAHA.111.044107}
\showDOI{\tempurl}


\bibitem[\protect\citeauthoryear{Lang, Badano, Mor-Avi, Afilalo, Armstrong,
  Ernande, Flachskampf, Foster, Goldstein, and Kuznetsova}{Lang
  et~al\mbox{.}}{2015}]%
        {RN47}
\bibfield{author}{\bibinfo{person}{Roberto~M Lang}, \bibinfo{person}{Luigi~P
  Badano}, \bibinfo{person}{Victor Mor-Avi}, \bibinfo{person}{Jonathan
  Afilalo}, \bibinfo{person}{Anderson Armstrong}, \bibinfo{person}{Laura
  Ernande}, \bibinfo{person}{Frank~A Flachskampf}, \bibinfo{person}{Elyse
  Foster}, \bibinfo{person}{Steven~A Goldstein}, {and} \bibinfo{person}{Tatiana
  Kuznetsova}.} \bibinfo{year}{2015}\natexlab{}.
\newblock \showarticletitle{Recommendations for cardiac chamber quantification
  by echocardiography in adults: an update from the American Society of
  Echocardiography and the European Association of Cardiovascular Imaging}.
\newblock \bibinfo{journal}{\emph{European Heart Journal-Cardiovascular
  Imaging}} \bibinfo{volume}{16}, \bibinfo{number}{3} (\bibinfo{year}{2015}),
  \bibinfo{pages}{233--271}.
\newblock
\showISSN{2047-2412}


\bibitem[\protect\citeauthoryear{Lara and Labrador}{Lara and Labrador}{2012}]%
        {RN72}
\bibfield{author}{\bibinfo{person}{Oscar~D Lara} {and}
  \bibinfo{person}{Miguel~A Labrador}.} \bibinfo{year}{2012}\natexlab{}.
\newblock \showarticletitle{A survey on human activity recognition using
  wearable sensors}.
\newblock \bibinfo{journal}{\emph{IEEE communications surveys \& tutorials}}
  \bibinfo{volume}{15}, \bibinfo{number}{3} (\bibinfo{year}{2012}),
  \bibinfo{pages}{1192--1209}.
\newblock
\showISSN{1553-877X}


\bibitem[\protect\citeauthoryear{Leng, San~Tan, Chai, Wang, Ghista, and
  Zhong}{Leng et~al\mbox{.}}{2015}]%
        {RN11}
\bibfield{author}{\bibinfo{person}{Shuang Leng}, \bibinfo{person}{Ru San~Tan},
  \bibinfo{person}{Kevin Tshun~Chuan Chai}, \bibinfo{person}{Chao Wang},
  \bibinfo{person}{Dhanjoo Ghista}, {and} \bibinfo{person}{Liang Zhong}.}
  \bibinfo{year}{2015}\natexlab{}.
\newblock \showarticletitle{The electronic stethoscope}.
\newblock \bibinfo{journal}{\emph{Biomedical engineering online}}
  \bibinfo{volume}{14}, \bibinfo{number}{1} (\bibinfo{year}{2015}),
  \bibinfo{pages}{66}.
\newblock
\showISSN{1475-925X}
\urldef\tempurl%
\url{https://doi.org/10.1186/s12938-015-0056-y}
\showDOI{\tempurl}


\bibitem[\protect\citeauthoryear{Levin, Dolgin, Fox, and Gorlin}{Levin
  et~al\mbox{.}}{1994}]%
        {RN7}
\bibfield{author}{\bibinfo{person}{R Levin}, \bibinfo{person}{M Dolgin},
  \bibinfo{person}{C Fox}, {and} \bibinfo{person}{R Gorlin}.}
  \bibinfo{year}{1994}\natexlab{}.
\newblock \showarticletitle{The Criteria Committee of the New York Heart
  Association: Nomenclature and Criteria for Diagnosis of Diseases of the Heart
  and Great Vessels}.
\newblock \bibinfo{journal}{\emph{LWW Handbooks}}  \bibinfo{volume}{9}
  (\bibinfo{year}{1994}), \bibinfo{pages}{344}.
\newblock


\bibitem[\protect\citeauthoryear{Li, Cummings, Lam, Graves, and Wu}{Li
  et~al\mbox{.}}{2009}]%
        {RN59}
\bibfield{author}{\bibinfo{person}{Changzhi Li}, \bibinfo{person}{Julie
  Cummings}, \bibinfo{person}{Jeffrey Lam}, \bibinfo{person}{Eric Graves},
  {and} \bibinfo{person}{Wenhsing Wu}.} \bibinfo{year}{2009}\natexlab{}.
\newblock \showarticletitle{Radar remote monitoring of vital signs}.
\newblock \bibinfo{journal}{\emph{IEEE Microwave Magazine}}
  \bibinfo{volume}{10}, \bibinfo{number}{1} (\bibinfo{year}{2009}),
  \bibinfo{pages}{47--56}.
\newblock
\showISSN{1527-3342}


\bibitem[\protect\citeauthoryear{Li, Rajagopalan, and Clifford}{Li
  et~al\mbox{.}}{2014}]%
        {RN54}
\bibfield{author}{\bibinfo{person}{Qiao Li}, \bibinfo{person}{Cadathur
  Rajagopalan}, {and} \bibinfo{person}{Gari~D Clifford}.}
  \bibinfo{year}{2014}\natexlab{}.
\newblock \showarticletitle{A machine learning approach to multi-level ECG
  signal quality classification}.
\newblock \bibinfo{journal}{\emph{Computer methods and programs in
  biomedicine}} \bibinfo{volume}{117}, \bibinfo{number}{3}
  (\bibinfo{year}{2014}), \bibinfo{pages}{435--447}.
\newblock
\showISSN{0169-2607}
\urldef\tempurl%
\url{https://doi.org/10.1016/j.cmpb.2014.09.002}
\showDOI{\tempurl}


\bibitem[\protect\citeauthoryear{Lip, Nieuwlaat, Pisters, Lane, and Crijns}{Lip
  et~al\mbox{.}}{2010}]%
        {RN15}
\bibfield{author}{\bibinfo{person}{Gregory~YH Lip}, \bibinfo{person}{Robby
  Nieuwlaat}, \bibinfo{person}{Ron Pisters}, \bibinfo{person}{Deirdre~A Lane},
  {and} \bibinfo{person}{Harry~JGM Crijns}.} \bibinfo{year}{2010}\natexlab{}.
\newblock \showarticletitle{Refining clinical risk stratification for
  predicting stroke and thromboembolism in atrial fibrillation using a novel
  risk factor-based approach: the euro heart survey on atrial fibrillation}.
\newblock \bibinfo{journal}{\emph{Chest}} \bibinfo{volume}{137},
  \bibinfo{number}{2} (\bibinfo{year}{2010}), \bibinfo{pages}{263--272}.
\newblock
\showISSN{0012-3692}


\bibitem[\protect\citeauthoryear{Lipton, Kale, and Wetzel}{Lipton
  et~al\mbox{.}}{2016}]%
        {RN109}
\bibfield{author}{\bibinfo{person}{Zachary~C Lipton}, \bibinfo{person}{David~C
  Kale}, {and} \bibinfo{person}{Randall Wetzel}.}
  \bibinfo{year}{2016}\natexlab{}.
\newblock \showarticletitle{Modeling missing data in clinical time series with
  rnns}.
\newblock \bibinfo{journal}{\emph{arXiv preprint arXiv:1606.04130}}
  (\bibinfo{year}{2016}).
\newblock


\bibitem[\protect\citeauthoryear{Liu, Li, Liu, and Wu}{Liu
  et~al\mbox{.}}{2019}]%
        {RN110}
\bibfield{author}{\bibinfo{person}{Yang Liu}, \bibinfo{person}{Zhenjiang Li},
  \bibinfo{person}{Zhidan Liu}, {and} \bibinfo{person}{Kaishun Wu}.}
  \bibinfo{year}{2019}\natexlab{}.
\newblock \showarticletitle{Real-time Arm Skeleton Tracking and Gesture
  Inference Tolerant to Missing Wearable Sensors}. In
  \bibinfo{booktitle}{\emph{Proceedings of the 17th Annual International
  Conference on Mobile Systems, Applications, and Services}}.
  \bibinfo{publisher}{ACM}, \bibinfo{pages}{287--299}.
\newblock
\showISBNx{1450366619}


\bibitem[\protect\citeauthoryear{Ma and Zhang}{Ma and Zhang}{2006}]%
        {RN88}
\bibfield{author}{\bibinfo{person}{T Ma} {and} \bibinfo{person}{Yuan-Ting
  Zhang}.} \bibinfo{year}{2006}\natexlab{}.
\newblock \showarticletitle{A correlation study on the variabilities in pulse
  transit time, blood pressure, and heart rate recorded simultaneously from
  healthy subjects}. In \bibinfo{booktitle}{\emph{2005 IEEE Engineering in
  Medicine and Biology 27th Annual Conference}}. \bibinfo{publisher}{IEEE},
  \bibinfo{pages}{996--999}.
\newblock
\showISBNx{0780387414}


\bibitem[\protect\citeauthoryear{Maddison, Rawstorn, Rolleston, Whittaker,
  Stewart, Benatar, Warren, Jiang, and Gant}{Maddison et~al\mbox{.}}{2014}]%
        {RN119}
\bibfield{author}{\bibinfo{person}{Ralph Maddison}, \bibinfo{person}{Jonathan~C
  Rawstorn}, \bibinfo{person}{Anna Rolleston}, \bibinfo{person}{Robyn
  Whittaker}, \bibinfo{person}{Ralph Stewart}, \bibinfo{person}{Jocelyne
  Benatar}, \bibinfo{person}{Ian Warren}, \bibinfo{person}{Yannan Jiang}, {and}
  \bibinfo{person}{Nicholas Gant}.} \bibinfo{year}{2014}\natexlab{}.
\newblock \showarticletitle{The remote exercise monitoring trial for
  exercise-based cardiac rehabilitation (REMOTE-CR): a randomised controlled
  trial protocol}.
\newblock \bibinfo{journal}{\emph{BMC Public Health}} \bibinfo{volume}{14},
  \bibinfo{number}{1} (\bibinfo{year}{2014}), \bibinfo{pages}{1236}.
\newblock
\showISSN{1471-2458}
\urldef\tempurl%
\url{https://doi.org/10.1186/1471-2458-14-1236}
\showDOI{\tempurl}


\bibitem[\protect\citeauthoryear{Mathews, Kambhamettu, and Barner}{Mathews
  et~al\mbox{.}}{2018}]%
        {RN113}
\bibfield{author}{\bibinfo{person}{Sherin~M Mathews}, \bibinfo{person}{Chandra
  Kambhamettu}, {and} \bibinfo{person}{Kenneth~E Barner}.}
  \bibinfo{year}{2018}\natexlab{}.
\newblock \showarticletitle{A novel application of deep learning for
  single-lead ECG classification}.
\newblock \bibinfo{journal}{\emph{Computers in biology and medicine}}
  \bibinfo{volume}{99} (\bibinfo{year}{2018}), \bibinfo{pages}{53--62}.
\newblock
\showISSN{0010-4825}
\urldef\tempurl%
\url{https://doi.org/10.1016/j.compbiomed.2018.05.013}
\showDOI{\tempurl}


\bibitem[\protect\citeauthoryear{McNamara, Wang, Herrin, Curtis, Bradley,
  Magid, Peterson, Blaney, Frederick, and Krumholz}{McNamara
  et~al\mbox{.}}{2006}]%
        {RN5}
\bibfield{author}{\bibinfo{person}{Robert~L McNamara}, \bibinfo{person}{Yongfei
  Wang}, \bibinfo{person}{Jeph Herrin}, \bibinfo{person}{Jeptha~P Curtis},
  \bibinfo{person}{Elizabeth~H Bradley}, \bibinfo{person}{David~J Magid},
  \bibinfo{person}{Eric~D Peterson}, \bibinfo{person}{Martha Blaney},
  \bibinfo{person}{Paul~D Frederick}, {and} \bibinfo{person}{Harlan~M
  Krumholz}.} \bibinfo{year}{2006}\natexlab{}.
\newblock \showarticletitle{Effect of door-to-balloon time on mortality in
  patients with ST-segment elevation myocardial infarction}.
\newblock \bibinfo{journal}{\emph{Journal of the American College of
  Cardiology}} \bibinfo{volume}{47}, \bibinfo{number}{11}
  (\bibinfo{year}{2006}), \bibinfo{pages}{2180--2186}.
\newblock
\showISSN{0735-1097}
\urldef\tempurl%
\url{https://doi.org/10.1016/j.jacc.2005.12.072}
\showDOI{\tempurl}


\bibitem[\protect\citeauthoryear{Members, Camm, Lip, De~Caterina, Savelieva,
  Atar, Hohnloser, Hindricks, Kirchhof, and Guidelines}{Members
  et~al\mbox{.}}{2012}]%
        {RN16}
\bibfield{author}{\bibinfo{person}{Authors/Task~Force Members},
  \bibinfo{person}{A~John Camm}, \bibinfo{person}{Gregory~YH Lip},
  \bibinfo{person}{Raffaele De~Caterina}, \bibinfo{person}{Irene Savelieva},
  \bibinfo{person}{Dan Atar}, \bibinfo{person}{Stefan~H Hohnloser},
  \bibinfo{person}{Gerhard Hindricks}, \bibinfo{person}{Paulus Kirchhof}, {and}
  \bibinfo{person}{ESC Committee for~Practice Guidelines}.}
  \bibinfo{year}{2012}\natexlab{}.
\newblock \showarticletitle{2012 focused update of the ESC guidelines for the
  management of atrial fibrillation: an update of the 2010 ESC guidelines for
  the management of atrial fibrillation developed with the special contribution
  of the European Heart Rhythm Association}.
\newblock \bibinfo{journal}{\emph{European heart journal}}
  \bibinfo{volume}{33}, \bibinfo{number}{21} (\bibinfo{year}{2012}),
  \bibinfo{pages}{2719--2747}.
\newblock
\showISSN{1522-9645}


\bibitem[\protect\citeauthoryear{Moody and Mark}{Moody and Mark}{2001}]%
        {RN57}
\bibfield{author}{\bibinfo{person}{George~B Moody} {and}
  \bibinfo{person}{Roger~G Mark}.} \bibinfo{year}{2001}\natexlab{}.
\newblock \showarticletitle{The impact of the MIT-BIH arrhythmia database}.
\newblock \bibinfo{journal}{\emph{IEEE Engineering in Medicine and Biology
  Magazine}} \bibinfo{volume}{20}, \bibinfo{number}{3} (\bibinfo{year}{2001}),
  \bibinfo{pages}{45--50}.
\newblock
\showISSN{0739-5175}
\urldef\tempurl%
\url{https://www.ncbi.nlm.nih.gov/pubmed/11446209}
\showURL{%
\tempurl}


\bibitem[\protect\citeauthoryear{Moore and Copel}{Moore and Copel}{2011}]%
        {RN48}
\bibfield{author}{\bibinfo{person}{Christopher~L Moore} {and}
  \bibinfo{person}{Joshua~A Copel}.} \bibinfo{year}{2011}\natexlab{}.
\newblock \showarticletitle{Point-of-care ultrasonography}.
\newblock \bibinfo{journal}{\emph{New England Journal of Medicine}}
  \bibinfo{volume}{364}, \bibinfo{number}{8} (\bibinfo{year}{2011}),
  \bibinfo{pages}{749--757}.
\newblock
\showISSN{0028-4793}
\urldef\tempurl%
\url{https://doi.org/10.1056/NEJMra0909487}
\showDOI{\tempurl}


\bibitem[\protect\citeauthoryear{Mortazavi, Nyamathi, Lee, Wilkerson,
  Ghasemzadeh, and Sarrafzadeh}{Mortazavi et~al\mbox{.}}{2013}]%
        {RN40}
\bibfield{author}{\bibinfo{person}{Bobak Mortazavi}, \bibinfo{person}{Suneil
  Nyamathi}, \bibinfo{person}{Sunghoon~Ivan Lee}, \bibinfo{person}{Thomas
  Wilkerson}, \bibinfo{person}{Hassan Ghasemzadeh}, {and}
  \bibinfo{person}{Majid Sarrafzadeh}.} \bibinfo{year}{2013}\natexlab{}.
\newblock \showarticletitle{Near-realistic mobile exergames with wireless
  wearable sensors}.
\newblock \bibinfo{journal}{\emph{IEEE Journal of Biomedical and Health
  Informatics}} \bibinfo{volume}{18}, \bibinfo{number}{2}
  (\bibinfo{year}{2013}), \bibinfo{pages}{449--456}.
\newblock
\showISSN{2168-2194}


\bibitem[\protect\citeauthoryear{Mortazavi, Pourhomayoun, Ghasemzadeh, Jafari,
  Roberts, and Sarrafzadeh}{Mortazavi et~al\mbox{.}}{2014b}]%
        {RN37}
\bibfield{author}{\bibinfo{person}{Bobak Mortazavi}, \bibinfo{person}{Mohammad
  Pourhomayoun}, \bibinfo{person}{Hassan Ghasemzadeh}, \bibinfo{person}{Roozbeh
  Jafari}, \bibinfo{person}{Christian~K Roberts}, {and} \bibinfo{person}{Majid
  Sarrafzadeh}.} \bibinfo{year}{2014}\natexlab{b}.
\newblock \showarticletitle{Context-aware data processing to enhance quality of
  measurements in wireless health systems: An application to met calculation of
  exergaming actions}.
\newblock \bibinfo{journal}{\emph{IEEE Internet of Things Journal}}
  \bibinfo{volume}{2}, \bibinfo{number}{1} (\bibinfo{year}{2014}),
  \bibinfo{pages}{84--93}.
\newblock
\showISSN{2327-4662}


\bibitem[\protect\citeauthoryear{Mortazavi, Downing, Bucholz, Dharmarajan,
  Manhapra, Li, Negahban, and Krumholz}{Mortazavi et~al\mbox{.}}{2016}]%
        {RN81}
\bibfield{author}{\bibinfo{person}{Bobak~J Mortazavi},
  \bibinfo{person}{Nicholas~S Downing}, \bibinfo{person}{Emily~M Bucholz},
  \bibinfo{person}{Kumar Dharmarajan}, \bibinfo{person}{Ajay Manhapra},
  \bibinfo{person}{Shu-Xia Li}, \bibinfo{person}{Sahand~N Negahban}, {and}
  \bibinfo{person}{Harlan~M Krumholz}.} \bibinfo{year}{2016}\natexlab{}.
\newblock \showarticletitle{Analysis of machine learning techniques for heart
  failure readmissions}.
\newblock \bibinfo{journal}{\emph{Circulation: Cardiovascular Quality and
  Outcomes}} \bibinfo{volume}{9}, \bibinfo{number}{6} (\bibinfo{year}{2016}),
  \bibinfo{pages}{629--640}.
\newblock
\showISSN{1941-7713}
\urldef\tempurl%
\url{https://doi.org/10.1161/CIRCOUTCOMES.116.003039}
\showDOI{\tempurl}


\bibitem[\protect\citeauthoryear{Mortazavi, Pourhomayoun, Alsheikh, Alshurafa,
  Lee, and Sarrafzadeh}{Mortazavi et~al\mbox{.}}{2014a}]%
        {RN103}
\bibfield{author}{\bibinfo{person}{Bobak~Jack Mortazavi},
  \bibinfo{person}{Mohammad Pourhomayoun}, \bibinfo{person}{Gabriel Alsheikh},
  \bibinfo{person}{Nabil Alshurafa}, \bibinfo{person}{Sunghoon~Ivan Lee}, {and}
  \bibinfo{person}{Majid Sarrafzadeh}.} \bibinfo{year}{2014}\natexlab{a}.
\newblock \showarticletitle{Determining the single best axis for exercise
  repetition recognition and counting on smartwatches}. In
  \bibinfo{booktitle}{\emph{2014 11th International Conference on Wearable and
  Implantable Body Sensor Networks}}. \bibinfo{publisher}{IEEE},
  \bibinfo{pages}{33--38}.
\newblock
\showISBNx{1479949590}


\bibitem[\protect\citeauthoryear{Nallamothu, Normand, Wang, Hofer, Brush~Jr,
  Messenger, Bradley, Rumsfeld, and Krumholz}{Nallamothu et~al\mbox{.}}{2015}]%
        {RN6}
\bibfield{author}{\bibinfo{person}{Brahmajee~K Nallamothu},
  \bibinfo{person}{Sharon-Lise~T Normand}, \bibinfo{person}{Yongfei Wang},
  \bibinfo{person}{Timothy~P Hofer}, \bibinfo{person}{John~E Brush~Jr},
  \bibinfo{person}{John~C Messenger}, \bibinfo{person}{Elizabeth~H Bradley},
  \bibinfo{person}{John~S Rumsfeld}, {and} \bibinfo{person}{Harlan~M
  Krumholz}.} \bibinfo{year}{2015}\natexlab{}.
\newblock \showarticletitle{Relation between door-to-balloon times and
  mortality after primary percutaneous coronary intervention over time: a
  retrospective study}.
\newblock \bibinfo{journal}{\emph{The Lancet}} \bibinfo{volume}{385},
  \bibinfo{number}{9973} (\bibinfo{year}{2015}), \bibinfo{pages}{1114--1122}.
\newblock
\showISSN{0140-6736}


\bibitem[\protect\citeauthoryear{Nathan, Paul, Prioleau, Niu, Mortazavi,
  Cambone, Veeraraghavan, Sabharwal, and Jafari}{Nathan et~al\mbox{.}}{2018}]%
        {RN38}
\bibfield{author}{\bibinfo{person}{Viswam Nathan}, \bibinfo{person}{Sudip
  Paul}, \bibinfo{person}{Temiloluwa Prioleau}, \bibinfo{person}{Li Niu},
  \bibinfo{person}{Bobak~J Mortazavi}, \bibinfo{person}{Stephen~A Cambone},
  \bibinfo{person}{Ashok Veeraraghavan}, \bibinfo{person}{Ashutosh Sabharwal},
  {and} \bibinfo{person}{Roozbeh Jafari}.} \bibinfo{year}{2018}\natexlab{}.
\newblock \showarticletitle{A Survey on Smart Homes for Aging in Place: Toward
  Solutions to the Specific Needs of the Elderly}.
\newblock \bibinfo{journal}{\emph{IEEE Signal Processing Magazine}}
  \bibinfo{volume}{35}, \bibinfo{number}{5} (\bibinfo{year}{2018}),
  \bibinfo{pages}{111--119}.
\newblock
\showISSN{1053-5888}


\bibitem[\protect\citeauthoryear{Nobel, Mayo, Hanley, Nadeau, and
  Daskalopoulou}{Nobel et~al\mbox{.}}{2014}]%
        {RN17}
\bibfield{author}{\bibinfo{person}{Lisa Nobel}, \bibinfo{person}{Nancy~E Mayo},
  \bibinfo{person}{James Hanley}, \bibinfo{person}{Lyne Nadeau}, {and}
  \bibinfo{person}{Stella~S Daskalopoulou}.} \bibinfo{year}{2014}\natexlab{}.
\newblock \showarticletitle{MyRisk\_Stroke calculator: a personalized stroke
  risk assessment tool for the general population}.
\newblock \bibinfo{journal}{\emph{Journal of Clinical Neurology}}
  \bibinfo{volume}{10}, \bibinfo{number}{1} (\bibinfo{year}{2014}),
  \bibinfo{pages}{1--9}.
\newblock
\showISSN{1738-6586}
\urldef\tempurl%
\url{https://doi.org/10.3988/jcn.2014.10.1.1}
\showDOI{\tempurl}


\bibitem[\protect\citeauthoryear{O'brien, Waeber, Parati, Staessen, and
  Myers}{O'brien et~al\mbox{.}}{2001}]%
        {RN51}
\bibfield{author}{\bibinfo{person}{Eoin O'brien}, \bibinfo{person}{Bernard
  Waeber}, \bibinfo{person}{Gianfranco Parati}, \bibinfo{person}{Jan Staessen},
  {and} \bibinfo{person}{Martin~G Myers}.} \bibinfo{year}{2001}\natexlab{}.
\newblock \showarticletitle{Blood pressure measuring devices: recommendations
  of the European Society of Hypertension}.
\newblock \bibinfo{journal}{\emph{Bmj}} \bibinfo{volume}{322},
  \bibinfo{number}{7285} (\bibinfo{year}{2001}), \bibinfo{pages}{531--536}.
\newblock
\showISSN{0959-8138}
\urldef\tempurl%
\url{https://doi.org/10.1136/bmj.322.7285.531}
\showDOI{\tempurl}


\bibitem[\protect\citeauthoryear{Olsen, Angell, Asma, Boutouyrie, Burger,
  Chirinos, Damasceno, Delles, Gimenez-Roqueplo, and Hering}{Olsen
  et~al\mbox{.}}{2016}]%
        {RN86}
\bibfield{author}{\bibinfo{person}{Michael~H Olsen}, \bibinfo{person}{Sonia~Y
  Angell}, \bibinfo{person}{Samira Asma}, \bibinfo{person}{Pierre Boutouyrie},
  \bibinfo{person}{Dylan Burger}, \bibinfo{person}{Julio~A Chirinos},
  \bibinfo{person}{Albertino Damasceno}, \bibinfo{person}{Christian Delles},
  \bibinfo{person}{Anne-Paule Gimenez-Roqueplo}, {and} \bibinfo{person}{Dagmara
  Hering}.} \bibinfo{year}{2016}\natexlab{}.
\newblock \showarticletitle{A call to action and a lifecourse strategy to
  address the global burden of raised blood pressure on current and future
  generations: the Lancet Commission on hypertension}.
\newblock \bibinfo{journal}{\emph{The Lancet}} \bibinfo{volume}{388},
  \bibinfo{number}{10060} (\bibinfo{year}{2016}), \bibinfo{pages}{2665--2712}.
\newblock
\showISSN{0140-6736}
\urldef\tempurl%
\url{https://doi.org/10.1016/S0140-6736(16)31134-5}
\showDOI{\tempurl}


\bibitem[\protect\citeauthoryear{Ong, Romano, Edgington, Aronow, Auerbach,
  Black, De~Marco, Escarce, Evangelista, and Hanna}{Ong et~al\mbox{.}}{2016}]%
        {RN33}
\bibfield{author}{\bibinfo{person}{Michael~K Ong}, \bibinfo{person}{Patrick~S
  Romano}, \bibinfo{person}{Sarah Edgington}, \bibinfo{person}{Harriet~U
  Aronow}, \bibinfo{person}{Andrew~D Auerbach}, \bibinfo{person}{Jeanne~T
  Black}, \bibinfo{person}{Teresa De~Marco}, \bibinfo{person}{Jose~J Escarce},
  \bibinfo{person}{Lorraine~S Evangelista}, {and} \bibinfo{person}{Barbara
  Hanna}.} \bibinfo{year}{2016}\natexlab{}.
\newblock \showarticletitle{Effectiveness of remote patient monitoring after
  discharge of hospitalized patients with heart failure: the better
  effectiveness after transition–heart failure (BEAT-HF) randomized clinical
  trial}.
\newblock \bibinfo{journal}{\emph{JAMA internal medicine}}
  \bibinfo{volume}{176}, \bibinfo{number}{3} (\bibinfo{year}{2016}),
  \bibinfo{pages}{310--318}.
\newblock
\showISSN{2168-6106}


\bibitem[\protect\citeauthoryear{Onusko}{Onusko}{2003}]%
        {RN67}
\bibfield{author}{\bibinfo{person}{Edward Onusko}.}
  \bibinfo{year}{2003}\natexlab{}.
\newblock \showarticletitle{Diagnosing secondary hypertension}.
\newblock \bibinfo{journal}{\emph{American family physician}}
  \bibinfo{volume}{67}, \bibinfo{number}{1} (\bibinfo{year}{2003}),
  \bibinfo{pages}{67--74}.
\newblock
\showISSN{0002-838X}
\urldef\tempurl%
\url{https://www.ncbi.nlm.nih.gov/pubmed/12537168}
\showURL{%
\tempurl}


\bibitem[\protect\citeauthoryear{{\O}stvik, Smistad, Aase, Haugen, and
  Lovstakken}{{\O}stvik et~al\mbox{.}}{2019}]%
        {RN44}
\bibfield{author}{\bibinfo{person}{Andreas {\O}stvik}, \bibinfo{person}{Erik
  Smistad}, \bibinfo{person}{Svein~Arne Aase}, \bibinfo{person}{Bj{\o}rn~Olav
  Haugen}, {and} \bibinfo{person}{Lasse Lovstakken}.}
  \bibinfo{year}{2019}\natexlab{}.
\newblock \showarticletitle{Real-time Standard View Classification in
  Transthoracic Echocardiography using Convolutional Neural Networks}.
\newblock \bibinfo{journal}{\emph{Ultrasound in medicine \& biology}}
  \bibinfo{volume}{45}, \bibinfo{number}{2} (\bibinfo{year}{2019}),
  \bibinfo{pages}{374--384}.
\newblock
\showISSN{0301-5629}


\bibitem[\protect\citeauthoryear{Ouchi and Doi}{Ouchi and Doi}{2013}]%
        {RN101}
\bibfield{author}{\bibinfo{person}{Kazushige Ouchi} {and}
  \bibinfo{person}{Miwako Doi}.} \bibinfo{year}{2013}\natexlab{}.
\newblock \showarticletitle{Smartphone-based monitoring system for activities
  of daily living for elderly people and their relatives etc}. In
  \bibinfo{booktitle}{\emph{Proceedings of the 2013 ACM conference on Pervasive
  and ubiquitous computing adjunct publication}}. \bibinfo{publisher}{ACM},
  \bibinfo{pages}{103--106}.
\newblock
\showISBNx{1450322158}


\bibitem[\protect\citeauthoryear{Palaniappan, Sundaraj, and Ahamed}{Palaniappan
  et~al\mbox{.}}{2013}]%
        {RN61}
\bibfield{author}{\bibinfo{person}{Rajkumar Palaniappan},
  \bibinfo{person}{Kenneth Sundaraj}, {and} \bibinfo{person}{Nizam~Uddin
  Ahamed}.} \bibinfo{year}{2013}\natexlab{}.
\newblock \showarticletitle{Machine learning in lung sound analysis: a
  systematic review}.
\newblock \bibinfo{journal}{\emph{Biocybernetics and Biomedical Engineering}}
  \bibinfo{volume}{33}, \bibinfo{number}{3} (\bibinfo{year}{2013}),
  \bibinfo{pages}{129--135}.
\newblock
\showISSN{0208-5216}


\bibitem[\protect\citeauthoryear{Pfeffer, Claggett, Assmann, Boineau, Anand,
  Clausell, Desai, Diaz, Fleg, and Gordeev}{Pfeffer et~al\mbox{.}}{2015}]%
        {RN124}
\bibfield{author}{\bibinfo{person}{Marc~A Pfeffer}, \bibinfo{person}{Brian
  Claggett}, \bibinfo{person}{Susan~F Assmann}, \bibinfo{person}{Robin
  Boineau}, \bibinfo{person}{Inder~S Anand}, \bibinfo{person}{Nadine Clausell},
  \bibinfo{person}{Akshay~S Desai}, \bibinfo{person}{Rafael Diaz},
  \bibinfo{person}{Jerome~L Fleg}, {and} \bibinfo{person}{Ivan Gordeev}.}
  \bibinfo{year}{2015}\natexlab{}.
\newblock \showarticletitle{Regional variation in patients and outcomes in the
  Treatment of Preserved Cardiac Function Heart Failure With an Aldosterone
  Antagonist (TOPCAT) trial}.
\newblock \bibinfo{journal}{\emph{Circulation}} \bibinfo{volume}{131},
  \bibinfo{number}{1} (\bibinfo{year}{2015}), \bibinfo{pages}{34--42}.
\newblock
\showISSN{0009-7322}
\urldef\tempurl%
\url{https://doi.org/10.1161/CIRCULATIONAHA.114.013255}
\showDOI{\tempurl}


\bibitem[\protect\citeauthoryear{Pires, Garcia, Pombo, and
  Fl{\'o}rez-Revuelta}{Pires et~al\mbox{.}}{2016}]%
        {RN102}
\bibfield{author}{\bibinfo{person}{Ivan Pires}, \bibinfo{person}{Nuno Garcia},
  \bibinfo{person}{Nuno Pombo}, {and} \bibinfo{person}{Francisco
  Fl{\'o}rez-Revuelta}.} \bibinfo{year}{2016}\natexlab{}.
\newblock \showarticletitle{From data acquisition to data fusion: a
  comprehensive review and a roadmap for the identification of activities of
  daily living using mobile devices}.
\newblock \bibinfo{journal}{\emph{Sensors}} \bibinfo{volume}{16},
  \bibinfo{number}{2} (\bibinfo{year}{2016}), \bibinfo{pages}{184}.
\newblock


\bibitem[\protect\citeauthoryear{Pisters, Lane, Nieuwlaat, De~Vos, Crijns, and
  Lip}{Pisters et~al\mbox{.}}{2010}]%
        {RN126}
\bibfield{author}{\bibinfo{person}{Ron Pisters}, \bibinfo{person}{Deirdre~A
  Lane}, \bibinfo{person}{Robby Nieuwlaat}, \bibinfo{person}{Cees~B De~Vos},
  \bibinfo{person}{Harry~JGM Crijns}, {and} \bibinfo{person}{Gregory~YH Lip}.}
  \bibinfo{year}{2010}\natexlab{}.
\newblock \showarticletitle{A novel user-friendly score (HAS-BLED) to assess
  1-year risk of major bleeding in patients with atrial fibrillation: the Euro
  Heart Survey}.
\newblock \bibinfo{journal}{\emph{Chest}} \bibinfo{volume}{138},
  \bibinfo{number}{5} (\bibinfo{year}{2010}), \bibinfo{pages}{1093--1100}.
\newblock
\showISSN{0012-3692}
\urldef\tempurl%
\url{https://doi.org/10.1378/chest.10-0134}
\showDOI{\tempurl}


\bibitem[\protect\citeauthoryear{Pitt, Pfeffer, Assmann, Boineau, Anand,
  Claggett, Clausell, Desai, Diaz, Fleg, et~al\mbox{.}}{Pitt
  et~al\mbox{.}}{2014}]%
        {pitt2014spironolactone}
\bibfield{author}{\bibinfo{person}{Bertram Pitt}, \bibinfo{person}{Marc~A
  Pfeffer}, \bibinfo{person}{Susan~F Assmann}, \bibinfo{person}{Robin Boineau},
  \bibinfo{person}{Inder~S Anand}, \bibinfo{person}{Brian Claggett},
  \bibinfo{person}{Nadine Clausell}, \bibinfo{person}{Akshay~S Desai},
  \bibinfo{person}{Rafael Diaz}, \bibinfo{person}{Jerome~L Fleg},
  {et~al\mbox{.}}} \bibinfo{year}{2014}\natexlab{}.
\newblock \showarticletitle{Spironolactone for heart failure with preserved
  ejection fraction}.
\newblock \bibinfo{journal}{\emph{New England Journal of Medicine}}
  \bibinfo{volume}{370}, \bibinfo{number}{15} (\bibinfo{year}{2014}),
  \bibinfo{pages}{1383--1392}.
\newblock


\bibitem[\protect\citeauthoryear{Pourhomayoun, Alshurafa, Mortazavi,
  Ghasemzadeh, Sideris, Sadeghi, Ong, Evangelista, Romano, and
  Auerbach}{Pourhomayoun et~al\mbox{.}}{2014}]%
        {RN115}
\bibfield{author}{\bibinfo{person}{Mohammad Pourhomayoun},
  \bibinfo{person}{Nabil Alshurafa}, \bibinfo{person}{Bobak Mortazavi},
  \bibinfo{person}{Hassan Ghasemzadeh}, \bibinfo{person}{Konstantinos Sideris},
  \bibinfo{person}{Bahman Sadeghi}, \bibinfo{person}{Michael Ong},
  \bibinfo{person}{Lorraine Evangelista}, \bibinfo{person}{Patrick Romano},
  {and} \bibinfo{person}{Andrew Auerbach}.} \bibinfo{year}{2014}\natexlab{}.
\newblock \showarticletitle{Multiple model analytics for adverse event
  prediction in remote health monitoring systems}. In
  \bibinfo{booktitle}{\emph{2014 IEEE Healthcare Innovation Conference (HIC)}}.
  \bibinfo{publisher}{IEEE}, \bibinfo{pages}{106--110}.
\newblock
\showISBNx{1467363642}


\bibitem[\protect\citeauthoryear{Reboussin, Allen, Griswold, Guallar, Hong,
  Lackland, Miller, Polonsky, Thompson-Paul, and Vupputuri}{Reboussin
  et~al\mbox{.}}{2018}]%
        {RN87}
\bibfield{author}{\bibinfo{person}{David~M Reboussin},
  \bibinfo{person}{Norrina~B Allen}, \bibinfo{person}{Michael~E Griswold},
  \bibinfo{person}{Eliseo Guallar}, \bibinfo{person}{Yuling Hong},
  \bibinfo{person}{Daniel~T Lackland}, \bibinfo{person}{Edgar Pete~R Miller},
  \bibinfo{person}{Tamar Polonsky}, \bibinfo{person}{Angela~M Thompson-Paul},
  {and} \bibinfo{person}{Suma Vupputuri}.} \bibinfo{year}{2018}\natexlab{}.
\newblock \showarticletitle{Systematic review for the 2017
  ACC/AHA/AAPA/ABC/ACPM/AGS/APhA/ASH/ASPC/NMA/PCNA guideline for the
  prevention, detection, evaluation, and management of high blood pressure in
  adults: a report of the American College of Cardiology/American Heart
  Association Task Force on Clinical Practice Guidelines}.
\newblock \bibinfo{journal}{\emph{Journal of the American College of
  Cardiology}} \bibinfo{volume}{71}, \bibinfo{number}{19}
  (\bibinfo{year}{2018}), \bibinfo{pages}{2176--2198}.
\newblock
\showISSN{0735-1097}


\bibitem[\protect\citeauthoryear{Ribeiro, Singh, and Guestrin}{Ribeiro
  et~al\mbox{.}}{2016}]%
        {RN128}
\bibfield{author}{\bibinfo{person}{Marco~Tulio Ribeiro},
  \bibinfo{person}{Sameer Singh}, {and} \bibinfo{person}{Carlos Guestrin}.}
  \bibinfo{year}{2016}\natexlab{}.
\newblock \showarticletitle{Why should I trust you?: Explaining the predictions
  of any classifier}. In \bibinfo{booktitle}{\emph{Proceedings of the 22nd ACM
  SIGKDD international conference on knowledge discovery and data mining}}.
  \bibinfo{publisher}{ACM}, \bibinfo{pages}{1135--1144}.
\newblock
\showISBNx{1450342329}


\bibitem[\protect\citeauthoryear{Rokni and Ghasemzadeh}{Rokni and
  Ghasemzadeh}{2016}]%
        {RN107}
\bibfield{author}{\bibinfo{person}{Seyed~Ali Rokni} {and}
  \bibinfo{person}{Hassan Ghasemzadeh}.} \bibinfo{year}{2016}\natexlab{}.
\newblock \showarticletitle{Plug-n-learn: automatic learning of computational
  algorithms in human-centered internet-of-things applications}. In
  \bibinfo{booktitle}{\emph{Proceedings of the 53rd Annual Design Automation
  Conference}}. \bibinfo{publisher}{ACM}, \bibinfo{pages}{139}.
\newblock
\showISBNx{1450342361}


\bibitem[\protect\citeauthoryear{Rokni and Ghasemzadeh}{Rokni and
  Ghasemzadeh}{2018}]%
        {RN108}
\bibfield{author}{\bibinfo{person}{Seyed~Ali Rokni} {and}
  \bibinfo{person}{Hassan Ghasemzadeh}.} \bibinfo{year}{2018}\natexlab{}.
\newblock \showarticletitle{Autonomous training of activity recognition
  algorithms in mobile sensors: A transfer learning approach in
  context-invariant views}.
\newblock \bibinfo{journal}{\emph{IEEE Transactions on Mobile Computing}}
  \bibinfo{volume}{17}, \bibinfo{number}{8} (\bibinfo{year}{2018}),
  \bibinfo{pages}{1764--1777}.
\newblock
\showISSN{1536-1233}


\bibitem[\protect\citeauthoryear{Rokni, Nourollahi, and Ghasemzadeh}{Rokni
  et~al\mbox{.}}{2018}]%
        {RN78}
\bibfield{author}{\bibinfo{person}{Seyed~Ali Rokni}, \bibinfo{person}{Marjan
  Nourollahi}, {and} \bibinfo{person}{Hassan Ghasemzadeh}.}
  \bibinfo{year}{2018}\natexlab{}.
\newblock \showarticletitle{Personalized human activity recognition using
  convolutional neural networks}. In \bibinfo{booktitle}{\emph{Thirty-Second
  AAAI Conference on Artificial Intelligence}}.
\newblock


\bibitem[\protect\citeauthoryear{Rymer and Rao}{Rymer and Rao}{2019}]%
        {RN122}
\bibfield{author}{\bibinfo{person}{Jennifer~A Rymer} {and}
  \bibinfo{person}{Sunil~V Rao}.} \bibinfo{year}{2019}\natexlab{}.
\newblock \showarticletitle{Enhancement of Risk Prediction With Machine
  Learning: Rise of the Machines}.
\newblock \bibinfo{journal}{\emph{JAMA network open}} \bibinfo{volume}{2},
  \bibinfo{number}{7} (\bibinfo{year}{2019}),
  \bibinfo{pages}{e196823--e196823}.
\newblock
\showISSN{2574-3805 (Electronic) 2574-3805 (Linking)}
\urldef\tempurl%
\url{https://doi.org/10.1001/jamanetworkopen.2019.6823}
\showDOI{\tempurl}


\bibitem[\protect\citeauthoryear{S{\'a}nchez-de-la Torre, Khalyfa,
  S{\'a}nchez-de-la Torre, Martinez-Alonso, Martinez-García, Barcel{\'o},
  Lloberes, Campos-Rodriguez, Capote, and Diaz-de Atauri}{S{\'a}nchez-de-la
  Torre et~al\mbox{.}}{2015}]%
        {RN114}
\bibfield{author}{\bibinfo{person}{Manuel S{\'a}nchez-de-la Torre},
  \bibinfo{person}{Abdelnaby Khalyfa}, \bibinfo{person}{Alicia
  S{\'a}nchez-de-la Torre}, \bibinfo{person}{Montserrat Martinez-Alonso},
  \bibinfo{person}{Miguel~{\'A}ngel Martinez-García}, \bibinfo{person}{Antonia
  Barcel{\'o}}, \bibinfo{person}{Patricia Lloberes}, \bibinfo{person}{Francisco
  Campos-Rodriguez}, \bibinfo{person}{Francisco Capote}, {and}
  \bibinfo{person}{Maria~Jos{\'e} Diaz-de Atauri}.}
  \bibinfo{year}{2015}\natexlab{}.
\newblock \showarticletitle{Precision medicine in patients with resistant
  hypertension and obstructive sleep apnea: blood pressure response to
  continuous positive airway pressure treatment}.
\newblock \bibinfo{journal}{\emph{Journal of the American College of
  Cardiology}} \bibinfo{volume}{66}, \bibinfo{number}{9}
  (\bibinfo{year}{2015}), \bibinfo{pages}{1023--1032}.
\newblock
\showISSN{0735-1097}
\urldef\tempurl%
\url{https://doi.org/10.1016/j.jacc.2015.06.1315}
\showDOI{\tempurl}


\bibitem[\protect\citeauthoryear{Sen, Subbaraju, Misra, Balan, and Lee}{Sen
  et~al\mbox{.}}{2015}]%
        {RN104}
\bibfield{author}{\bibinfo{person}{Sougata Sen}, \bibinfo{person}{Vigneshwaran
  Subbaraju}, \bibinfo{person}{Archan Misra}, \bibinfo{person}{Rajesh~Krishna
  Balan}, {and} \bibinfo{person}{Youngki Lee}.}
  \bibinfo{year}{2015}\natexlab{}.
\newblock \showarticletitle{The case for smartwatch-based diet monitoring}. In
  \bibinfo{booktitle}{\emph{2015 IEEE international conference on pervasive
  computing and communication workshops (PerCom workshops)}}.
  \bibinfo{publisher}{IEEE}, \bibinfo{pages}{585--590}.
\newblock
\showISBNx{1479984256}


\bibitem[\protect\citeauthoryear{Sinharay, Ghosh, Deshpande, Alam, Banerjee,
  and Pal}{Sinharay et~al\mbox{.}}{2016}]%
        {RN64}
\bibfield{author}{\bibinfo{person}{Arijit Sinharay}, \bibinfo{person}{Deb
  Ghosh}, \bibinfo{person}{Parijat Deshpande}, \bibinfo{person}{Shahnawaz
  Alam}, \bibinfo{person}{Rohan Banerjee}, {and} \bibinfo{person}{Arpan Pal}.}
  \bibinfo{year}{2016}\natexlab{}.
\newblock \showarticletitle{Smartphone Based Digital Stethoscope for Connected
  Health--A Direct Acoustic Coupling Technique}. In
  \bibinfo{booktitle}{\emph{2016 IEEE First International Conference on
  Connected Health: Applications, Systems and Engineering Technologies
  (CHASE)}}. \bibinfo{publisher}{IEEE}, \bibinfo{pages}{193--198}.
\newblock
\showISBNx{1509009434}


\bibitem[\protect\citeauthoryear{Solis, Pakbin, Akbari, Mortazavi, and
  Jafari}{Solis et~al\mbox{.}}{2019}]%
        {RN133}
\bibfield{author}{\bibinfo{person}{Roger Solis}, \bibinfo{person}{Arash
  Pakbin}, \bibinfo{person}{Ali Akbari}, \bibinfo{person}{Bobak~J Mortazavi},
  {and} \bibinfo{person}{Roozbeh Jafari}.} \bibinfo{year}{2019}\natexlab{}.
\newblock \showarticletitle{A human-centered wearable sensing platform with
  intelligent automated data annotation capabilities}. In
  \bibinfo{booktitle}{\emph{Proceedings of the International Conference on
  Internet of Things Design and Implementation}}. \bibinfo{publisher}{ACM},
  \bibinfo{pages}{255--260}.
\newblock
\showISBNx{1450362834}


\bibitem[\protect\citeauthoryear{Song, Rajan, Thiagarajan, and Spanias}{Song
  et~al\mbox{.}}{2018}]%
        {RN121}
\bibfield{author}{\bibinfo{person}{Huan Song}, \bibinfo{person}{Deepta Rajan},
  \bibinfo{person}{Jayaraman~J Thiagarajan}, {and} \bibinfo{person}{Andreas
  Spanias}.} \bibinfo{year}{2018}\natexlab{}.
\newblock \showarticletitle{Attend and diagnose: Clinical time series analysis
  using attention models}. In \bibinfo{booktitle}{\emph{Thirty-Second AAAI
  Conference on Artificial Intelligence}}.
\newblock


\bibitem[\protect\citeauthoryear{Sriram, Shin, Choudhury, and Kotz}{Sriram
  et~al\mbox{.}}{2009}]%
        {RN53}
\bibfield{author}{\bibinfo{person}{Janani~C Sriram}, \bibinfo{person}{Minho
  Shin}, \bibinfo{person}{Tanzeem Choudhury}, {and} \bibinfo{person}{David
  Kotz}.} \bibinfo{year}{2009}\natexlab{}.
\newblock \showarticletitle{Activity-aware ECG-based patient authentication for
  remote health monitoring}. In \bibinfo{booktitle}{\emph{Proceedings of the
  2009 international conference on Multimodal interfaces}}.
  \bibinfo{publisher}{ACM}, \bibinfo{pages}{297--304}.
\newblock
\showISBNx{1605587729}


\bibitem[\protect\citeauthoryear{Steinhubl, Waalen, Edwards, Ariniello, Mehta,
  Ebner, Carter, Baca-Motes, Felicione, and Sarich}{Steinhubl
  et~al\mbox{.}}{2018}]%
        {RN55}
\bibfield{author}{\bibinfo{person}{Steven~R Steinhubl}, \bibinfo{person}{Jill
  Waalen}, \bibinfo{person}{Alison~M Edwards}, \bibinfo{person}{Lauren~M
  Ariniello}, \bibinfo{person}{Rajesh~R Mehta}, \bibinfo{person}{Gail~S Ebner},
  \bibinfo{person}{Chureen Carter}, \bibinfo{person}{Katie Baca-Motes},
  \bibinfo{person}{Elise Felicione}, {and} \bibinfo{person}{Troy Sarich}.}
  \bibinfo{year}{2018}\natexlab{}.
\newblock \showarticletitle{Effect of a home-based wearable continuous ECG
  monitoring patch on detection of undiagnosed atrial fibrillation: the mSToPS
  randomized clinical trial}.
\newblock \bibinfo{journal}{\emph{Jama}} \bibinfo{volume}{320},
  \bibinfo{number}{2} (\bibinfo{year}{2018}), \bibinfo{pages}{146--155}.
\newblock
\showISSN{0098-7484}
\urldef\tempurl%
\url{https://doi.org/10.1001/jama.2018.8102}
\showDOI{\tempurl}


\bibitem[\protect\citeauthoryear{Suresh, Gong, and Guttag}{Suresh
  et~al\mbox{.}}{2018}]%
        {RN130}
\bibfield{author}{\bibinfo{person}{Harini Suresh}, \bibinfo{person}{Jen~J
  Gong}, {and} \bibinfo{person}{John~V Guttag}.}
  \bibinfo{year}{2018}\natexlab{}.
\newblock \showarticletitle{Learning tasks for multitask learning: Heterogenous
  patient populations in the icu}. In \bibinfo{booktitle}{\emph{Proceedings of
  the 24th ACM SIGKDD International Conference on Knowledge Discovery \& Data
  Mining}}. \bibinfo{publisher}{ACM}, \bibinfo{pages}{802--810}.
\newblock
\showISBNx{1450355528}


\bibitem[\protect\citeauthoryear{Thomas, Nathan, Zong, Akinbola, Aroul,
  Philipose, Soundarapandian, Shi, and Jafari}{Thomas et~al\mbox{.}}{2014}]%
        {RN89}
\bibfield{author}{\bibinfo{person}{Simi~Susan Thomas}, \bibinfo{person}{Viswam
  Nathan}, \bibinfo{person}{Chengzhi Zong}, \bibinfo{person}{Ebunoluwa
  Akinbola}, \bibinfo{person}{Antoine Lourdes~Praveen Aroul},
  \bibinfo{person}{Lijoy Philipose}, \bibinfo{person}{Karthikeyan
  Soundarapandian}, \bibinfo{person}{Xiangrong Shi}, {and}
  \bibinfo{person}{Roozbeh Jafari}.} \bibinfo{year}{2014}\natexlab{}.
\newblock \showarticletitle{BioWatch-A wrist watch based signal acquisition
  system for physiological signals including blood pressure}. In
  \bibinfo{booktitle}{\emph{2014 36th Annual International Conference of the
  IEEE Engineering in Medicine and Biology Society}}.
  \bibinfo{publisher}{IEEE}, \bibinfo{pages}{2286--2289}.
\newblock
\showISBNx{1424479290}


\bibitem[\protect\citeauthoryear{Trujillo, Nathan, Cot{\'e}, and
  Jafari}{Trujillo et~al\mbox{.}}{2017}]%
        {RN92}
\bibfield{author}{\bibinfo{person}{Zachary Trujillo}, \bibinfo{person}{Viswam
  Nathan}, \bibinfo{person}{Gerard~L Cot{\'e}}, {and} \bibinfo{person}{Roozbeh
  Jafari}.} \bibinfo{year}{2017}\natexlab{}.
\newblock \showarticletitle{Design and parametric analysis of a wearable
  dual-photoplethysmograph based system for pulse wave velocity detection}. In
  \bibinfo{booktitle}{\emph{2017 IEEE International Symposium on Circuits and
  Systems (ISCAS)}}. \bibinfo{publisher}{IEEE}, \bibinfo{pages}{1--4}.
\newblock
\showISBNx{1467368539}


\bibitem[\protect\citeauthoryear{Van~Laerhoven, Borazio, and
  Burdinski}{Van~Laerhoven et~al\mbox{.}}{2015}]%
        {RN105}
\bibfield{author}{\bibinfo{person}{Kristof Van~Laerhoven},
  \bibinfo{person}{Marko Borazio}, {and} \bibinfo{person}{Jan~Hendrik
  Burdinski}.} \bibinfo{year}{2015}\natexlab{}.
\newblock \showarticletitle{Wear is your mobile? Investigating phone carrying
  and use habits with a wearable device}.
\newblock \bibinfo{journal}{\emph{Frontiers in ICT}}  \bibinfo{volume}{2}
  (\bibinfo{year}{2015}), \bibinfo{pages}{10}.
\newblock
\showISSN{2297-198X}


\bibitem[\protect\citeauthoryear{Vasan, Beiser, Seshadri, Larson, Kannel,
  D'Agostino, and Levy}{Vasan et~al\mbox{.}}{2002}]%
        {RN65}
\bibfield{author}{\bibinfo{person}{Ramachandran~S Vasan},
  \bibinfo{person}{Alexa Beiser}, \bibinfo{person}{Sudha Seshadri},
  \bibinfo{person}{Martin~G Larson}, \bibinfo{person}{William~B Kannel},
  \bibinfo{person}{Ralph~B D'Agostino}, {and} \bibinfo{person}{Daniel Levy}.}
  \bibinfo{year}{2002}\natexlab{}.
\newblock \showarticletitle{Residual lifetime risk for developing hypertension
  in middle-aged women and men: The Framingham Heart Study}.
\newblock \bibinfo{journal}{\emph{Jama}} \bibinfo{volume}{287},
  \bibinfo{number}{8} (\bibinfo{year}{2002}), \bibinfo{pages}{1003--1010}.
\newblock
\showISSN{0098-7484}
\urldef\tempurl%
\url{https://doi.org/10.1001/jama.287.8.1003}
\showDOI{\tempurl}


\bibitem[\protect\citeauthoryear{Vinci, Lindner, Barbon, Mann, Hofmann, Duda,
  Weigel, and Koelpin}{Vinci et~al\mbox{.}}{2013}]%
        {RN60}
\bibfield{author}{\bibinfo{person}{Gabor Vinci}, \bibinfo{person}{Stefan
  Lindner}, \bibinfo{person}{Francesco Barbon}, \bibinfo{person}{Sebastian
  Mann}, \bibinfo{person}{Maximilian Hofmann}, \bibinfo{person}{Alexander
  Duda}, \bibinfo{person}{Robert Weigel}, {and} \bibinfo{person}{Alexander
  Koelpin}.} \bibinfo{year}{2013}\natexlab{}.
\newblock \showarticletitle{Six-port radar sensor for remote respiration rate
  and heartbeat vital-sign monitoring}.
\newblock \bibinfo{journal}{\emph{IEEE Transactions on Microwave Theory and
  Techniques}} \bibinfo{volume}{61}, \bibinfo{number}{5}
  (\bibinfo{year}{2013}), \bibinfo{pages}{2093--2100}.
\newblock
\showISSN{0018-9480}


\bibitem[\protect\citeauthoryear{Wang, Chen, Hao, Peng, and Hu}{Wang
  et~al\mbox{.}}{2019}]%
        {RN77}
\bibfield{author}{\bibinfo{person}{Jindong Wang}, \bibinfo{person}{Yiqiang
  Chen}, \bibinfo{person}{Shuji Hao}, \bibinfo{person}{Xiaohui Peng}, {and}
  \bibinfo{person}{Lisha Hu}.} \bibinfo{year}{2019}\natexlab{}.
\newblock \showarticletitle{Deep learning for sensor-based activity
  recognition: A survey}.
\newblock \bibinfo{journal}{\emph{Pattern Recognition Letters}}
  \bibinfo{volume}{119} (\bibinfo{year}{2019}), \bibinfo{pages}{3--11}.
\newblock
\showISSN{0167-8655}


\bibitem[\protect\citeauthoryear{Wasserlauf, You, Patel, Valys, Albert, and
  Passman}{Wasserlauf et~al\mbox{.}}{2019}]%
        {RN85}
\bibfield{author}{\bibinfo{person}{Jeremiah Wasserlauf}, \bibinfo{person}{Cindy
  You}, \bibinfo{person}{Ruchi Patel}, \bibinfo{person}{Alexander Valys},
  \bibinfo{person}{David Albert}, {and} \bibinfo{person}{Rod Passman}.}
  \bibinfo{year}{2019}\natexlab{}.
\newblock \showarticletitle{Smartwatch Performance for the Detection and
  Quantification of Atrial Fibrillation}.
\newblock \bibinfo{journal}{\emph{Circulation: Arrhythmia and
  Electrophysiology}} \bibinfo{volume}{12}, \bibinfo{number}{6}
  (\bibinfo{year}{2019}), \bibinfo{pages}{e006834}.
\newblock
\showISSN{1941-3084}
\urldef\tempurl%
\url{https://doi.org/10.1161/CIRCEP.118.006834}
\showDOI{\tempurl}


\bibitem[\protect\citeauthoryear{Weyer, Menden, Leicht, Leonhardt, and
  Wartzek}{Weyer et~al\mbox{.}}{2015}]%
        {RN42}
\bibfield{author}{\bibinfo{person}{S{\"o}ren Weyer}, \bibinfo{person}{Tobias
  Menden}, \bibinfo{person}{Lennart Leicht}, \bibinfo{person}{Steffen
  Leonhardt}, {and} \bibinfo{person}{Tobias Wartzek}.}
  \bibinfo{year}{2015}\natexlab{}.
\newblock \showarticletitle{Development of a wearable multi-frequency impedance
  cardiography device}.
\newblock \bibinfo{journal}{\emph{Journal of medical engineering \&
  technology}} \bibinfo{volume}{39}, \bibinfo{number}{2}
  (\bibinfo{year}{2015}), \bibinfo{pages}{131--137}.
\newblock
\showISSN{0309-1902}
\urldef\tempurl%
\url{https://doi.org/10.3109/03091902.2014.990161}
\showDOI{\tempurl}


\bibitem[\protect\citeauthoryear{Wright, Verouhis, Gamble, Swedberg, Sharpe,
  and Doughty}{Wright et~al\mbox{.}}{2003}]%
        {RN69}
\bibfield{author}{\bibinfo{person}{SP Wright}, \bibinfo{person}{D Verouhis},
  \bibinfo{person}{G Gamble}, \bibinfo{person}{K Swedberg}, \bibinfo{person}{N
  Sharpe}, {and} \bibinfo{person}{RN Doughty}.}
  \bibinfo{year}{2003}\natexlab{}.
\newblock \showarticletitle{Factors influencing the length of hospital stay of
  patients with heart failure}.
\newblock \bibinfo{journal}{\emph{European Journal of Heart Failure}}
  \bibinfo{volume}{5}, \bibinfo{number}{2} (\bibinfo{year}{2003}),
  \bibinfo{pages}{201--209}.
\newblock
\showISSN{1388-9842}
\urldef\tempurl%
\url{https://doi.org/10.1016/s1388-9842(02)00201-5}
\showDOI{\tempurl}


\bibitem[\protect\citeauthoryear{Xie, Jean, Burke, Lobell, and Ermon}{Xie
  et~al\mbox{.}}{2016}]%
        {RN106}
\bibfield{author}{\bibinfo{person}{Michael Xie}, \bibinfo{person}{Neal Jean},
  \bibinfo{person}{Marshall Burke}, \bibinfo{person}{David Lobell}, {and}
  \bibinfo{person}{Stefano Ermon}.} \bibinfo{year}{2016}\natexlab{}.
\newblock \showarticletitle{Transfer learning from deep features for remote
  sensing and poverty mapping}. In \bibinfo{booktitle}{\emph{Thirtieth AAAI
  Conference on Artificial Intelligence}}.
\newblock


\bibitem[\protect\citeauthoryear{Yao, Weaver, Langley, George, and Hardin}{Yao
  et~al\mbox{.}}{2017}]%
        {RN99}
\bibfield{author}{\bibinfo{person}{Jianchu Yao}, \bibinfo{person}{Elizabeth~MB
  Weaver}, \bibinfo{person}{Brandon~D Langley}, \bibinfo{person}{Stephanie~M
  George}, {and} \bibinfo{person}{Sonya~R Hardin}.}
  \bibinfo{year}{2017}\natexlab{}.
\newblock \showarticletitle{Monitoring peripheral edema of heart failure
  patients at home: Device, algorithm, and clinic study}. In
  \bibinfo{booktitle}{\emph{2017 39th Annual International Conference of the
  IEEE Engineering in Medicine and Biology Society (EMBC)}}.
  \bibinfo{publisher}{IEEE}, \bibinfo{pages}{4074--4077}.
\newblock
\showISBNx{1509028099}


\bibitem[\protect\citeauthoryear{Yildirim}{Yildirim}{2018}]%
        {RN56}
\bibfield{author}{\bibinfo{person}{{\"O}zal Yildirim}.}
  \bibinfo{year}{2018}\natexlab{}.
\newblock \showarticletitle{A novel wavelet sequence based on deep
  bidirectional LSTM network model for ECG signal classification}.
\newblock \bibinfo{journal}{\emph{Computers in biology and medicine}}
  \bibinfo{volume}{96} (\bibinfo{year}{2018}), \bibinfo{pages}{189--202}.
\newblock
\showISSN{0010-4825}
\urldef\tempurl%
\url{https://doi.org/10.1016/j.compbiomed.2018.03.016}
\showDOI{\tempurl}


\bibitem[\protect\citeauthoryear{Yu, Wang, Chau, Chan, Kong, Tang, Christensen,
  Stadler, and Lau}{Yu et~al\mbox{.}}{2005}]%
        {RN43}
\bibfield{author}{\bibinfo{person}{Cheuk-Man Yu}, \bibinfo{person}{LI Wang},
  \bibinfo{person}{Elaine Chau}, \bibinfo{person}{Raymond Hon-Wah Chan},
  \bibinfo{person}{Shun-Ling Kong}, \bibinfo{person}{Man-Oi Tang},
  \bibinfo{person}{Jill Christensen}, \bibinfo{person}{Robert~W Stadler}, {and}
  \bibinfo{person}{Chu-Pak Lau}.} \bibinfo{year}{2005}\natexlab{}.
\newblock \showarticletitle{Intrathoracic impedance monitoring in patients with
  heart failure: correlation with fluid status and feasibility of early warning
  preceding hospitalization}.
\newblock \bibinfo{journal}{\emph{Circulation}} \bibinfo{volume}{112},
  \bibinfo{number}{6} (\bibinfo{year}{2005}), \bibinfo{pages}{841--848}.
\newblock
\showISSN{0009-7322}
\urldef\tempurl%
\url{https://doi.org/10.1161/CIRCULATIONAHA.104.492207}
\showDOI{\tempurl}


\bibitem[\protect\citeauthoryear{Y\"ur\"ur, Liu, and Moreno}{Y\"ur\"ur
  et~al\mbox{.}}{2015}]%
        {RN100}
\bibfield{author}{\bibinfo{person}{\"Ozg\"ur Y\"ur\"ur},
  \bibinfo{person}{Chi~Harold Liu}, {and} \bibinfo{person}{Wilfrido Moreno}.}
  \bibinfo{year}{2015}\natexlab{}.
\newblock \showarticletitle{Light-weight online unsupervised posture detection
  by smartphone accelerometer}.
\newblock \bibinfo{journal}{\emph{IEEE Internet of Things Journal}}
  \bibinfo{volume}{2}, \bibinfo{number}{4} (\bibinfo{year}{2015}),
  \bibinfo{pages}{329--339}.
\newblock
\showISSN{2327-4662}


\bibitem[\protect\citeauthoryear{Zhang, Poon, Chan, Tsang, and Wu}{Zhang
  et~al\mbox{.}}{2006}]%
        {RN116}
\bibfield{author}{\bibinfo{person}{Yuan-ting Zhang}, \bibinfo{person}{Carmen~CY
  Poon}, \bibinfo{person}{Chun-hung Chan}, \bibinfo{person}{Martin~WW Tsang},
  {and} \bibinfo{person}{Kin-fai Wu}.} \bibinfo{year}{2006}\natexlab{}.
\newblock \showarticletitle{A health-shirt using e-textile materials for the
  continuous and cuffless monitoring of arterial blood pressure}. In
  \bibinfo{booktitle}{\emph{2006 3rd IEEE/EMBS International Summer School on
  Medical Devices and Biosensors}}. \bibinfo{publisher}{IEEE},
  \bibinfo{pages}{86--89}.
\newblock
\showISBNx{078039786X}


\bibitem[\protect\citeauthoryear{Zhang}{Zhang}{2015}]%
        {RN84}
\bibfield{author}{\bibinfo{person}{Zhilin Zhang}.}
  \bibinfo{year}{2015}\natexlab{}.
\newblock \showarticletitle{Photoplethysmography-based heart rate monitoring in
  physical activities via joint sparse spectrum reconstruction}.
\newblock \bibinfo{journal}{\emph{IEEE transactions on biomedical engineering}}
  \bibinfo{volume}{62}, \bibinfo{number}{8} (\bibinfo{year}{2015}),
  \bibinfo{pages}{1902--1910}.
\newblock
\showISSN{0018-9294}
\urldef\tempurl%
\url{https://doi.org/10.1109/TBME.2015.2406332}
\showDOI{\tempurl}


\bibitem[\protect\citeauthoryear{Ziaeian and Fonarow}{Ziaeian and
  Fonarow}{2016}]%
        {RN9}
\bibfield{author}{\bibinfo{person}{Boback Ziaeian} {and}
  \bibinfo{person}{Gregg~C Fonarow}.} \bibinfo{year}{2016}\natexlab{}.
\newblock \showarticletitle{Epidemiology and aetiology of heart failure}.
\newblock \bibinfo{journal}{\emph{Nature Reviews Cardiology}}
  \bibinfo{volume}{13}, \bibinfo{number}{6} (\bibinfo{year}{2016}),
  \bibinfo{pages}{368}.
\newblock
\showISSN{1759-5010}
\urldef\tempurl%
\url{https://doi.org/10.1038/nrcardio.2016.25}
\showDOI{\tempurl}


\end{thebibliography}

\appendix

\section{Normal Cardiac Physiology}
\label{sec:appendix}
In order to better understand the opportunities discussed in this work, we provide a brief medical review of normal function of the heart. 

The heart consists of four chambers- two atria and two ventricles, one of each composing the right heart and one of each composing the left heart. The two atria (singular atrium) drive blood into the two ventricles, and the ventricles drive blood forward. The right heart takes deoxygenated blood from the body and pumps it through the lungs. The left heart takes blood from the lungs and pumps it through the body. The left heart does more work than the right heart, and the pressures that it produces are higher than the pressures produced by the right heart. There are four valves in the heart. The right heart valves are the tricuspid and the pulmonic. The tricuspid valve is between the right atrium and ventricle, the pulmonic valve is between the right ventricle and the lungs. The valves of the left heart are the mitral and the aortic. The mitral valve is between the left atrium and ventricle, while the aortic valve is between the left ventricle and the rest of the body.

The heart cycles through two phases as it beats- systole and diastole. In systole, the heart contracts and drives blood forward. In diastole, the heart relaxes and refills with blood. One common metric of cardiac function is blood pressure, represented by two numbers- a systolic and a diastolic pressure. The systolic pressure is the pressure in the arteries as blood is actively pumped out of the heart. The diastolic pressure is the pressure in the arteries as the heart relaxes. The heart typically spends about one third of the time in systole, and two thirds of the time in diastole. The normal heart produces two sounds, called S1 and S2. S1 is caused by the closing of the mitral and tricuspid valves at the beginning of systole. Their closure prevents the backwards flow of blood from the ventricles into the atria. S2 is caused by the closing of the aortic and pulmonic valves. Their closure prevents the backwards flow of blood from the body or lungs backwards into the heart.
\end{document}